\documentclass[a4paper,fleqn,usenatbib]{mnras}

\usepackage[T1]{fontenc}
\usepackage{ae,aecompl}

\usepackage{graphicx}	
\usepackage{amsmath}	
\usepackage{amssymb}	
\usepackage{xpatch}

\usepackage{hyperref}
\hypersetup{
	colorlinks=true,
	breaklinks=true,
	allcolors=blue,
	frenchlinks=true
}

\makeatletter
\xpatchcmd\NAT@citex
{%
	\@citea\NAT@hyper@{%
		\NAT@nmfmt{\NAT@nm}%
		\hyper@natlinkbreak{\NAT@aysep\NAT@spacechar}{\@citeb\@extra@b@citeb}%
		\NAT@date
	}%
}
{%
	\@citea
	\NAT@nmfmt{\NAT@nm}%
	\NAT@aysep\NAT@spacechar
	\NAT@hyper@{\NAT@date}%
}
{}{}
\xpatchcmd\NAT@citex
{%
	\@citea\NAT@hyper@{%
		\NAT@nmfmt{\NAT@nm}%
		\hyper@natlinkbreak{\NAT@spacechar\NAT@@open\if*#1*\else#1\NAT@spacechar\fi}%
		{\@citeb\@extra@b@citeb}%
		\NAT@date
	}%
}
{
	\@citea
	\NAT@nmfmt{\NAT@nm}%
	\NAT@spacechar\NAT@@open\if*#1*\else#1\NAT@spacechar\fi
	\NAT@hyper@{\NAT@date}%
}
{}{}
\makeatother

\usepackage[normalem]{ulem} 
\usepackage{amsbsy}
\usepackage{mathrsfs,bm}
\newcommand{\bcdot}{\ensuremath{%
  \mathchoice%
   {\mskip\thinmuskip\lower0.2ex\hbox{\scalebox{1.5}{$\cdot$}}\mskip\thinmuskip}}%
   {\mskip\thinmuskip\lower0.2ex\hbox{\scalebox{1.5}{$\cdot$}}\mskip\thinmuskip}%
   {\lower0.3ex\hbox{\scalebox{1.2}{$\cdot$}}}%
   {\lower0.3ex\hbox{\scalebox{1.2}{$\cdot$}}}%
}

\newcommand{\btimes}{\ensuremath{\boldsymbol{\times}}}
\newcommand{\bnabla}{\ensuremath{\boldsymbol{\nabla}}}
\newcommand{\vect}[1]{\boldsymbol{#1}}
\newcommand{\vecbf}[1]{\mathbfit{#1}}

\newcommand\Tstrut{\rule{0pt}{2.6ex}}         
\newcommand\Bstrut{\rule[-0.9ex]{0pt}{0pt}}   

\newcommand{\pjet}{L_\mathrm{jet}}

\newcommand{\mdot}{\dot{M}}

\newcommand{\tcool}{t_\mathrm{cool}}
\newcommand{\arepo}{\textsc{arepo}}
\newcommand{\weinberger}{Weinberger et al., in prep.}
\newcommand{\halpha}{H$\alpha$ }

\usepackage                             {hyperref}
\usepackage[dvipsnames]{xcolor}
\usepackage{array}

\title[AGN feedback of light jets]
      {Self-regulated AGN feedback of light jets in cool-core galaxy clusters}

\author[K. Ehlert et al.]{
K. Ehlert$^{1,2}$\thanks{E-mail: kehlert@aip.de},
R. Weinberger$^{3}$,
C. Pfrommer$^{1}$,
R. Pakmor$^{4}$,
V. Springel$^{4}$
\\
$^{1}$Leibniz Institute for Astrophysics, An der Sternwarte 16, D-14482 Potsdam, Germany\\
$^{2}$Institut f\"ur Physik und Astronomie, Universit\"at Potsdam, Karl-Liebknecht-Str. 24/25, 14476 Golm, Germany\\
$^{3}$Canadian Institute for Theoretical Astrophysics, 60 St. George Street, Toronto, ON M5S 3H8, Canada\\
$^{4}$Max-Planck-Institut f\"ur Astrophysik, Karl-Schwarzschild-Str. 1, D-85741 Garching, Germany\\
}
\date{Accepted XXX. Received YYY; in original form ZZZ}

\pubyear{2022}

\begin{document}
\label{firstpage}
\pagerange{\pageref{firstpage}--\pageref{lastpage}}
\maketitle

\begin{abstract}
  Heating from active galactic nuclei (AGN) is thought to stabilize cool-core
  clusters, limiting star formation and cooling flows. We employ radiative
  magneto-hydrodynamic (MHD) simulations to model light AGN jet feedback with
  different accretion modes (Bondi-Hoyle-Lyttleton and cold accretion) in an
  idealised Perseus-like cluster. Independent of the probed accretion model,
  accretion efficiency, jet density and resolution, the cluster self-regulates
  with central entropies and cooling times consistent with observed cool-core
  clusters in this non-cosmological setting. We find that increased jet
  efficiencies lead to more intermittent jet powers and enhanced star formation
  rates. Our fiducial low-density jets can easily be deflected by orbiting cold
  gaseous filaments, which redistributes angular momentum and leads to more
  extended cold gas distributions and isotropic bubble distributions. In
  comparison to our fiducial low momentum-density jets, high momentum-density
  jets heat less efficiently and enable the formation of a persistent cold-gas
  disc perpendicular to the jets that is centrally confined. Cavity luminosities
  measured from our simulations generally reflect the cooling luminosities of
  the intracluster medium (ICM) and correspond to averaged jet powers that are
  relatively insensitive to short periods of low-luminosity jet injection. Cold
  gas structures in our MHD simulations with low momentum-density jets generally
  show a variety of morphologies ranging from discy to very extended filamentary
  structures. In particular, magnetic fields are crucial to inhibit the
  formation of unrealistically massive cold gas discs by redistributing angular
  momentum between the hot and cold phases and by fostering the formation of
  elongated cold filaments that are supported by magnetic pressure.
\end{abstract}




\section{Introduction}
Cool-core (CC) clusters with central cooling times smaller than 1~Gyr form a subclass of galaxy clusters. However, the expected cooling flows are absent. Instead these clusters posses low star formation rates and low central entropies \citep{Peterson2006}. Jets driven by the central AGNs inflate buoyantly rising bubbles that are observed as X-ray cavities. The mechanical luminosity of AGNs estimated from cavity enthalpy appears to be tightly linked to the cooling luminosity \cite[e.g.,][]{Birzan2004,Rafferty2006,Diehl2008}, leaving sufficient heating energy to offset the energy losses by the cooling ICM and  establishing feedback from AGNs as the main heating source \citep{McNamara2012,Fabian2012} in CC clusters.

The AGN is fueled by cooling gas accreted by the central super massive black hole (SMBH). The exact accretion mechanism remains uncertain. In many numerical simulations, Bondi-Hoyle-Lyttleton accretion \citep{Bondi1952,Hoyle1941} is employed due to its simplicity. However, observations find that Bondi accretion provides insufficient power to fuel active jets in clusters \citep[e.g.,][]{Cavagnolo2011,McNamara2011,Russell2015,Fujita2016a,Russell2018a}. The ICM is prone to the thermal instability, which  acts when the cooling time $t_\mathrm{cool}$ is of order or shorter than the free fall time $t_\mathrm{ff}$ \citep{Mccourt2012}. Cold gas may condense in the galaxy or can become thermally unstable as it is dragged up by the AGN (e.g., \citealt{McNamara2016,Russell2017a,Tremblay2018}, but see, \citealt{Jones2017}). The condensing cold gas is then predicted to rain on the central SMBH as \textit{cold accretion} \citep{Pizzolato2005,Sharma2012,Gaspari2017b}. \cite{Tremblay2016a} report observational evidence of such a cold clumpy accretion flow towards a SMBH. In many CC clusters, cold gas takes a filamentary shape \citep[e.g.,][]{Russell2019,Olivares2019} where clusters with low values of $t_\mathrm{cool}/t_\mathrm{ff}$ show more massive filaments (e.g., \citealt{Cavagnolo2008,Voit2015c,Lakhchaura2018b} but see \citealt{Martz2020a}).

Hydrodynamic simulations of galaxy clusters are able to produce a self-regulated feedback loop using a cold gas triggered SMBH accretion model which is coupled to the AGN feedback injection \citep[e.g.,][]{Gaspari2012,Li2014b,Prasad2015}. However, there are various AGN jet models proposed in the literature that range from high-momentum density jets \citep[with or without precession,][which heat mainly via mixing the cold and hot phases]{Sternberg2008,Hillel2014} to low-momentum density jets of various ICM-to-jet density ratios \citep{Yang2016,Weinberger2017,Beckmann2022}, which can be more easily deflected by dense gas clouds in the path of propagation or by coherent and/or turbulent motions of the ambient ICM \citep{Heinz2006}.

When including star formation, \cite{Li2015} find a substantial suppression of the star formation rate compared to the expectation from an unmediated cooling flow for large enough efficiencies. A too low feedback efficiency, however, leads to high star formation rates $\gg100\,\mathrm{M}_\odot\,\mathrm{yr}^{-1}$, inconsistent with observations \citep[e.g.,][]{Fogarty2015}. Cold gas properties are highly interconnected with the heating-cooling cycle: AGN-induced uplift is key in shaping the spatial distribution of cold gas \citep{Yang2016a}, producing a wide range of cold gas cloud morphologies. Jets tend to shatter these structures \citep{Beckmann2019a}. The addition of magnetic fields in simulations leads to a suppression of unrealistically massive cold gas discs that tend to appear in purely hydrodynamic simulations \citep{Wang2021}.

We simulate hot, low-density jets in an idealised magnetized CC cluster. In a companion paper (\weinberger), we compare our low-density jet implementation to other AGN feedback implementations and study uncertainties arising due to resolution, parameter and model choices in detail. Here, we focus on (i) self-regulated feedback in CC clusters, and its dependence on the accretion prescription and jet properties, (ii) studying how mechanical X-ray cavity powers are related to the cooling luminosities and (simulated) jet luminosities, (iii) addressing the relevance of magnetic fields in redistributing the AGN feedback energy and in shaping the cold gas kinematics, and (iv) exploring how sensitive these results are to jet and accretion parameter choices.

The outline of our work is as follows. In Section~\ref{sec:methods}, we describe our initial conditions and simulation setup. In Section~\ref{sec:selfreg} we demonstrate that independent of model choices, i.e.\ the adopted accretion model, probed jet efficiency and jet density, we obtain a self-regulated CC cluster. We analyse mechanical luminosities derived from X-ray cavities of our runs in Section~\ref{sec:cavities} and study magnetic fields and the emerging cold gas in Section~\ref{sec:bfield_coldgas}. We discuss our results in Section~\ref{sec:discussion} and conclude in Section~\ref{sec:conclusions}.

\section{Methods and simulation models}
\label{sec:methods}
We compute three-dimensional MHD simulations of AGN feedback in an idealised Perseus-like cluster with the moving-mesh code \arepo\ \citep{Springel2010, Pakmor2016}. We use an HLLD Riemann solver \citep{Pakmor2011,Pakmor2013} with the Powell 8-wave scheme for magnetic divergence control \citep{Powell1999}. This solver has been shown to match several observed properties of magnetic fields in galaxies \citep{Pakmor2017,2018MNRAS.481.4410P} and the circumgalactic medium \citep{2020MNRAS.498.3125P}. Moreover, the growth rates and saturation properties of the small-scale dynamo, the kinetic and magnetic power spectra, and statistics of magnetic curvature forces are in line with expectations from idealized turbulent box simulations of magnetic dynamos \citep{2022MNRAS.515.4229P}, thus providing circumstantial evidence that our employed numerical method of moving mesh magneto-hydrodynamics with Powell cleaning delivers accurate results. In the following we describe the initial conditions, our ISM modeling, and further details of our jet/accretion models and simulation runs.

\begin{figure*}
	\centering
	\includegraphics[trim=0cm 0cm 0cm 0cm,clip=true, width=1\textwidth]{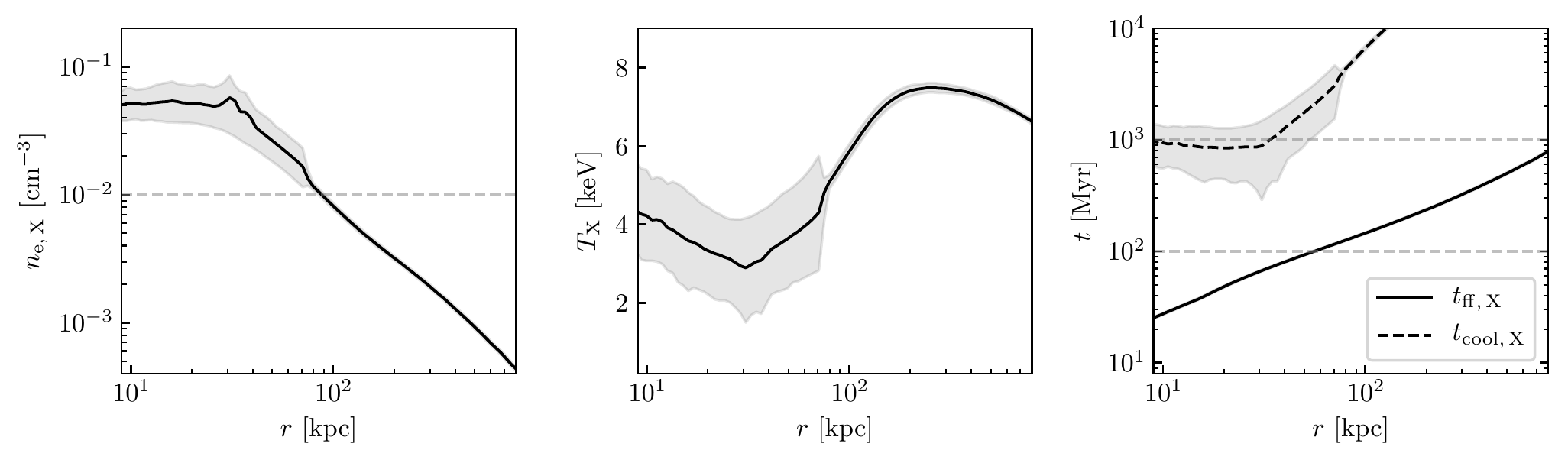}
	\caption{Radial profiles of the X-ray luminosity weighted electron number density (left), temperature (center), cooling time $t_\mathrm{cool}$ and free fall time $t_\mathrm{ff}$ (right) in our initial conditions. The shaded regions indicate the 10th and 90th percentiles. Only gas cells with $0.2\,\mathrm{keV}<k_\rmn{B}T<10\,\mathrm{keV}$ are considered.}
	\label{fig:timescales}
\end{figure*}

\subsection{Initial conditions}
We adopt the radial electron density profile from \cite{Churazov2003} rescaled to a cosmology with $h=0.67$:
\begin{equation}
\begin{split}
n_\mathrm{e} &= 46\times10^{-3}\left[1+\left(\frac{r}{60\ \text{kpc}}\right)^2\right]^{-1.8}\text{cm}^{-3}\\
&\quad +4.7\times10^{-3}\left[1+\left(\frac{r}{210\ \text{kpc}}\right)^2\right]^{-0.87}\text{cm}^{-3},
\label{eq:ne}
\end{split}
\end{equation}
which is set up in hydrostatic equilibrium with the gravitational potential comprised of an NFW cluster potential with virial radius $R_{200,\mathrm{NFW}}=2\,\mathrm{Mpc}$, mass $M_{200,\mathrm{NFW}}=8\times10^{14}\,\mathrm{M}_\odot$ and concentration parameter 5. On top of this cluster potential, we add a central galaxy potential approximated as an isothermal sphere with $M_{200,\mathrm{ISO}}=2.4\times10^{11}\,\mathrm{M}_\odot$ with $R_{200,\mathrm{ISO}}=15\,\mathrm{kpc}$ \citep{Mathews2006}. We only account for the gravity that results from a static background potential, neglecting effects of self-gravity. We confirmed that it has negligible effect in comparison runs.

The turbulent magnetic field is introduced as a Gaussian random field with a Kolmogorov slope on scales smaller than $k_\mathrm{inj}=37.5^{-1}\,\mathrm{kpc}^{-1}$ and white noise on larger scales. The chosen scale is consistent with observations \citep{Vacca2018}. We use a constant thermal-to-magnetic pressure ratio of $X_{B,\mathrm{ICM}}=P_B/P_\rmn{th}=0.0125$, lower than in previous work to reduce rather large resulting values of Faraday rotation measure \citep{Ehlert2021}. The initial mesh consists of three nested meshes, with increasing resolution towards the center. We used an iterative procedure to provide an initial magnetic field that obeys the magnetic divergence constraint, $\bnabla\bcdot\vecbf{B}=0$. Details of the divergence-free field setup can be found in \cite{Ehlert2018}.

We introduce temperature fluctuations in the initial condition, seeding thermal instabilities at various stages to prevent the sudden emergence of a large amount of cold gas. We multiply the hydrostatic temperatures by values drawn from a Gaussian random field for $\delta T/T$ with dispersion $\sigma=2$ and mean $\mu=1$. The power spectrum of temperature fluctuations follows the Kolmogorov slope on scales smaller than $k_\mathrm{inj}=37.5^{-1}\,\mathrm{kpc}^{-1}$ and corresponds to white noise on larger scales, consistent with the power spectrum of the fluctuations in the  magnetic field strength.

In addition, we seed velocity fluctuations in the central $800\,\mathrm{kpc}$ of the initial conditions to mimic orbiting substructures and random gas motions that result from gravitational potential rearrangements in the virialisation process. We initialize individual field components as a Gaussian random field with standard deviation $\sigma=70\,\mathrm{km}\,\mathrm{s}^{-1}$. Powers on other scales are set to zero. The history of the initial velocity field is quickly erased as the jet significantly perturbs the velocity field within the crucial inner $200\,\mathrm{kpc}$. Note, we use the same velocity fluctuations for all simulations analysed to simplify comparison across the different simulations.

In order to ease comparison to X-ray observations, we weight some quantities with the cooling luminosity of the X-ray emitting gas cells. Here, we only include cells with temperatures $0.2\,\mathrm{keV}<k_\rmn{B}T<10\,\mathrm{keV}$. The cooling luminosity is directly taken from the simulation. We refer to this weighting scheme as ``X-ray luminosity weighted'' in the following. In Fig.~\ref{fig:timescales}, we show radial profiles of the X-ray luminosity weighted electron number density, temperature, cooling time $t_\mathrm{cool}$ and free fall time $t_\mathrm{ff}$.

The free-fall time $t_\mathrm{ff}$ \citep{Sharma2012} (or \textit{local dynamical time}, \citealt{Mccourt2012}) is given by
\begin{equation}
t_\mathrm{ff}=\sqrt{2r/g}
\end{equation}
where $g=\mathrm{d}\Phi/\mathrm{d}r$ corresponds to the local acceleration due to gravity. The cooling time is defined via
\begin{equation}
t_\mathrm{cool}=\frac{\varepsilon_\rmn{th}}{\dot\varepsilon_\rmn{cool}},
\end{equation}
where $\varepsilon_\rmn{th}$ is the thermal energy density and $\dot\varepsilon_\rmn{cool}$ is the cooling luminosity for X-ray emitting gas (with $0.2\,\mathrm{keV}<k_\rmn{B}T<10\,\mathrm{keV}$) in our simulations. Figure~\ref{fig:timescales} shows that the Gaussian temperature fluctuations introduce cells that start condensing soon after the start of the simulations while more than 90 per cent of the gas has cooling times exceeding 300 Myrs.

\subsection{Cooling and star formation}
The modeling of cooling, star formation and stellar feedback follows \citet{Vogelsberger2013} with updates and parameter choices consistent with the IllustrisTNG model \citep{Pillepich2018}.
Cooling is modeled down to temperatures of $10^4$~K using primordial and metal-line cooling assuming a constant metallicity of 0.3 times the solar value, which is motivated by an observed uniform ICM iron abundance $Z_\mathrm{Fe}\approx0.3$ in solar units \citep{Werner2013a} for $r\lesssim1.5\,\mathrm{Mpc}$. Treatment of star formation and supernova feedback are part of the ISM model described in \cite{Springel2003}. The model implements star formation as a stochastic process, where the star formation probability is tied to the free fall time of the gas and calibrated to follow the observed Kennicutt relation \citep{KennicuttJr.1998}. Supernovae feedback heats the hot ICM and evaporates cold clouds. This leads to a tightly regulated regime for star formation. In the model this translates to an ISM that is pressurized by star formation feedback such that cold and hot phase coexist in pressure equilibrium, with the pressure given by a density dependent effective equation of state (eEOS). Above a density threshold of $n_\mathrm{e}=0.13\,\mathrm{cm}^{-3}$, gas can exist either on this effective equation of state (see red line in Fig.~\ref{fig:phasediagram}) and form stars or as non star-forming phase with larger temperatures, but not below it.
Gas on this eEOS is by definition star forming and, as mentioned in the previous subsection, also the main fuel for SMBH accretion.
Stellar feedback beyond that implicitly accounted for by the eEOS is not modeled. At the chosen halo mass scale these stellar feedback effects are anyhow subdominant.

\subsection{Jet}
\label{sec:jet}
Provided the AGN cavities are in pressure equilibrium with the ambient ICM and supported by (entrained) thermal plasma \citep{Croston2014,Croston2018}, the X-ray surface brightness maps imply a low cavity density and constrain this gas to be much hotter than the surrounding ICM. Because the bright X-ray emission of the ICM along the line of sight is projected onto the potentially faint emission from the cavities, X-ray spectroscopy alone is unable to probe the existence of very hot diffuse thermal plasma in excess of tens of keV filling the cavities. In fact, the high cavity-to-ICM density contrast was used to constrain the temperature of thermal plasma potentially supporting the cavities to $k_\rmn{B}T > 20$--50 keV \citep{Nulsen2002,Blanton2003,Sanders2007}.

Instead, a thermally supported cavity provides a Sunyaev-Zel'dovich (SZ) signal that is distinguishable from the signal of a cavity supported by magnetic fields and non-thermal relativistic particles, which themselves contribute minimally to the SZ effect \citep{Pfrommer2005,Ehlert2019}. Observations of the Sunyaev-Zel'dovich Effect in MS 0735.6+7421 show a clear deficit in signal consistent with temperatures of $\gtrsim1000\,\mathrm{keV}$ in the case of thermal pressure supported bubbles \citep{Abdulla2018}. This corresponds to an ICM-to-jet density contrast of $\rho_\mathrm{ICM}/\rho_\mathrm{jet}>10^2$ to $2\times10^2$ (and likely much larger). We therefore adopt $\rho_\mathrm{jet}=10^{-28}\mathrm{g}\,\mathrm{cm}^{-3}$ (i.e., $\rho_\mathrm{ICM}/\rho_\mathrm{jet}\sim3\times10^3$--$10^4$) as our fiducial jet density. Jets are injected in a bi-directional fashion with zero opening angle from a spherical region with radius $r=r_\mathrm{acc}/3$ (\weinberger). We inject a helical magnetic field in the jet fluid with a magnetic-to-thermal pressure ratio $X_{B,\mathrm{jet}}$.

\cite{Talbot2021} model accretion of geometrically thin discs and launch jets using the \citet{Blandford1977} model. They find that the jet direction varies mildly by 25$^\circ$ over 10~Myr for the most extreme case, when the jet is launched into the accretion disc, while it remains stable otherwise. Lower black hole-spin values lead to a more efficient reorientation of the spin \citep{Beckmann2019a}, because only little accreted material (with a different spin) is needed to torque the SMBH spin while maximally spinning Kerr black holes require the accretion of at least the mass of the black hole itself to change the spin orientation by unity. Realistic accretion models require the modeling of a geometrically thick disc in the low-Eddington accretion regime as expected for jets. This implies turbulent discs in which the mageto-rotational instability can transport angular momentum to larger radii, thus limiting the accreted angular momentum and black-hole spin reorientation, which has a direct consequence on the jet launching direction. Importantly, chaotic cold accretion implies the feeding of the accretion disc with material of random angular momentum, such that the average angular momentum does not appreciably change over time scales of 1~Gyr, which justifies our choice of a steady jet direction. Similar theoretical arguments have been made by \citet{Nixon2013}, suggesting that rapidly reorienting jets would be an indication against a Blandford-Znajek jet launching mechanism.

To set up the jet state, we select gas cells within a spherical region with radius $r=r_\mathrm{acc}/3$ and set their density to the pre-defined value $\rho_\mathrm{jet}$. Unlike in previous studies \citep{Weinberger2017}, this is done directly in the center and the mass that is removed from (or added to) this jet region is not redistributed to the surroundings, but added to the gravitating mass of the black hole, ensuring total mass conservation. This mass is treated as a reservoir for future accreted gas in the black hole accretion routine. The total energy in the system is reduced by $\Delta m \left<u\right>$, where $\Delta m$ is the mass change and $\left<u\right>$ is the ambient specific thermal energy. The gas cells in the spherical shell outside the jet region but within a radius of $r_\mathrm{acc}$ are used to determine these ambient gas properties. In the jet region, we add internal energy to ensure that cells are at least in pressure equilibrium with the surroundings. Note that we do not allow for internal energy to be reduced at this step.
From the jet energy available at a timestep, $L_\mathrm{jet}\mathrm{d}t$, we subtract the energy required to set up the jet state with its fixed density and in pressure equilibrium with its surroundings and inject the remaining energy in the form of kinetic energy bidirectionally without opening angle into the jet region. To trace the jet, we initialize a passive scalar $X_\mathrm{jet}=1$ in the jet region and advect it with the fluid. Our target mass is based on distance from the centre as
\begin{equation}
m_\mathrm{target}=m_{\mathrm{target},0}\exp(r/\mathrm{100}\,\mathrm{kpc}),
\end{equation}
with cells at the outskirts limited to a maximum volume with a cell radius $r_\rmn{cell}=370\,\mathrm{kpc}$. To sustain the strong density gradient between jet and ICM, we additionally refine cells with $X_\mathrm{jet}>10^{-3}$ to a target volume $V_\mathrm{jet,target}$, where we limit the volume ratio between neighboring cells to 4 \citep[see][ and \weinberger\  for further details on the jet implementation]{Weinberger2017}. Table~\ref{table:sims} shows the corresponding values for the discussed parameters in our runs.

\subsection{Accretion}
One focus of this study is to analyse the impact of the employed accretion model. Here, we use Bondi accretion and chaotic cold accretion with implementation details presented in the following. For the accretion rate estimate we use gas properties from cells within radius $r_\mathrm{acc}$ from the SMBH (excluding the jet region). We assume an initial black hole mass of $4\times10^9\,\mathrm{M}_\odot$, in rough agreement with expectations from black hole-halo scaling relations. This is an order of magnitude more massive than the SMBH in NGC1275 in the center of the Perseus cluster \citep{Wilman2004,Scharwachter2013}, which is an outlier in the black hole - host scaling relations \citep{Sani2018}.

\subsubsection{Cold accretion}
We parameterize the SMBH accretion rate $\mdot$ in the cold mode as
\begin{equation}
\mdot_\rmn{cold}=\epsilon\frac{M_\mathrm{cold}}{t_\mathrm{ff}},
\end{equation}
where $M_\mathrm{cold}$ includes star forming gas and gas with a temperature below $2\times10^4\,\mathrm{K}$. Note, $M_\mathrm{cold}$ corresponds to the total gas mass of the cell. We do not explicitly compute the mass component of the cold phase in the ISM model. We then drain mass $\Delta M_{\mathrm{bh},i} = \dot{M}_{\mathrm{cold},i}\Delta t$ from all cold gas cells $i$ during timestep $\Delta t$ and increment the black hole mass by the total drained gas mass $\Sigma_i\Delta M_{\mathrm{bh},i}$. When calculating the free-fall time $t_\mathrm{ff}$, we only consider gravitational acceleration due to the static galaxy and cluster potential.

Not all gas arrives at the BH within $t_\mathrm{ff}$ due to the angular momentum of the gas. Processes that lead to angular momentum cancellation, e.g.\ cloud collisions \citep{Gaspari2017b}, are not necessarily efficient enough under all circumstances. In addition, unresolved small-scale feedback may evaporate a fractions of the cold gas.
To take these effects into account, we introduce the parameter $\epsilon<1$ which represents a simple parametrization of the importance of these effects.
Due to the high pressure environment of a massive galaxy cluster, our temperature threshold is below the effective temperature of most cold gas given by the effective equation of state of the interstellar medium \citep[ISM,][see Fig.~\ref{fig:phasediagram}]{Springel2003}. Therefore the SMBH is de-facto mostly accreting gas from the star forming phase.

The jet power $L_\mathrm{jet}$ is proportional to a fraction $\eta$ of the accreted rest-mass energy
\begin{equation}
\pjet=\eta\mdot_\rmn{cold} c^2=\eta\epsilon\frac{M_\mathrm{cold}}{t_\mathrm{ff}} c^2,
\end{equation}
where $c$ denotes the speed of light.
\subsubsection{Bondi accretion}
We compare the cold gas based accretion model to the frequently used Bondi accretion estimate. The Bondi accretion rate is given by
\begin{equation}
	\dot{M}_\mathrm{bondi}=\frac{4\pi G^2M^2\rho}{c_\rmn{s}^3},
\end{equation}
where $M$ is the SMBH mass, $c_\rmn{s}$ is the speed of sound, $\rho$ is the mass density of the accreting medium. Assuming spherical accretion, the accretion rate must be below the Eddington limit. In practice, this limit is never reached in our runs. Analogously to cold accretion, the jet power is given by
\begin{equation}
	\pjet=\eta\dot{M}_\mathrm{bondi} c^2.
\end{equation}

\begin{table*}
	\begin{center}
		\begin{tabular}{c c c c c c c c c c c c c}
			\hline
			Label & accretion & $\epsilon$ & $\eta$ & $\rho_\mathrm{jet}$ & $m_{\mathrm{target},0}$ & $V_\mathrm{jet,target}^{1/3}$ & $r_\mathrm{acc}$ & radiative cooling & $X_{B,\mathrm{ICM}}$ & $X_{B,\mathrm{jet}}$ \Tstrut\Bstrut\\
			& & & & $[\mathrm{g}\,\mathrm{cm}^{-3}]$ & $[10^5\,\mathrm{M}_\odot]$ & $[\mathrm{kpc}]$ & $[\mathrm{kpc}]$ \Bstrut\\
			\hline\Tstrut\Bstrut
			\texttt{Fiducial} & cc & 0.1 & 0.01 & $10^{-28}$ & 15 & 0.65 & 2 & $\checkmark$ & 0.0125 & 0.1\\
			\texttt{Dense}& cc & 0.1 & 0.01 & $10^{-25}$ & 15 & 0.65 & 2 & $\checkmark$ & 0.0125 & 0.1\\
			\texttt{HD} & cc & 0.1 & 0.01 & $10^{-28}$ & 15 & 0.65 & 2 & $\checkmark$ & 0 & 0\\
			\texttt{Bondi} & bo & - & 0.01 & $10^{-28}$ & 15 & 0.65 & 2 & $\checkmark$ & 0.0125 & 0.1\\
			\texttt{NoBH}& - & - & - & - & 15 & - & - & $\checkmark$ & 0.0125 & -\\
			\texttt{NoBHNoCool}& - & - & - & - & 15 & - & - & - & 0.0125 & -\\
			\\
			\texttt{HR} & cc & 0.1 & 0.01 & $10^{-28}$ & 1.5 & 0.3 & 1 & $\checkmark$ & 0.0125 & 0.1\\
			\texttt{HDHR} & cc & 0.1 & 0.01 & $10^{-28}$ & 1.5 & 0.3 & 1 & $\checkmark$ & 0 & 0\\
			\texttt{BondiHR} & bo & - & 0.01 & $10^{-28}$ & 1.5 & 0.3 & 1 & $\checkmark$ & 0.0125 & 0.1\\
			\\
			Varying accretion\\
                        parameters\\
			\hline
			& cc & 0.1 & 0.1 & $10^{-28}$ & 15 & 0.65 & 2 & $\checkmark$ & 0.0125 & 0.1\\
			& cc & 0.01 & 0.01 & $10^{-28}$ & 15 & 0.65 & 2 & $\checkmark$ & 0.0125 & 0.1\\
			& cc & 0.01 & 0.001 & $10^{-28}$ & 15 & 0.65 & 2 & $\checkmark$ & 0.0125 & 0.1\\
			& cc & 1 & 0.0001 & $10^{-28}$ & 15 & 0.65 & 2 & $\checkmark$ & 0.0125 & 0.1\\
			& bo & - & 0.001 & $10^{-28}$ & 15 & 0.65 & 2 & $\checkmark$ & 0.0125 & 0.1\\
			\hline

		\end{tabular}
	\end{center}
	\caption{Simulation parameters with accretion model (bo for Bondi accretion, and cc for chaotic cold accretion), efficiency parameters $\epsilon$ and $\eta$, jet density $\rho_\mathrm{jet}$, target mass $m_\mathrm{target,0}$, jet target volume $V_\mathrm{jet,target}$, radius of accretion region $r_\mathrm{acc}$, thermal-to-magnetic pressure ratio in the ICM $X_{B,\mathrm{ICM}}$ and in the jet $X_{B,\mathrm{jet}}$. }
	\label{table:sims}
\end{table*}

\subsection{Simulation runs}
Simulations discussed in the following are summarized in Table~\ref{table:sims}.
The \texttt{Fiducial} model uses cold accretion with our fiducial efficiency parameters $\epsilon=0.1$ and $\eta=0.01$. To study the impact of magnetic fields, we rerun \texttt{Fiducial} by excluding MHD, which we call \texttt{HD}. Our \texttt{Bondi} run with fiducial Bondi accretion allows comparisons with a different mode of accretion. We analyse effects of numerical resolution by rerunning \texttt{Fiducial}, \texttt{HD} and \texttt{Bondi} at 10 times higher mass resolution, reduced accretion radius $r_\mathrm{acc}=1\,\mathrm{kpc}$ and a factor of 2 higher spatial resolution in the jet (see Table~\ref{table:sims}). We refer to the high resolution runs as \texttt{HR}, \texttt{HDHR} and \texttt{BondiHR}, respectively. We also vary the efficiency parameters $\epsilon$ and $\eta$ of the cold accretion model and increase the jet density by $10^3$ in our \texttt{Dense} model. To assess the effects of our jet feedback, we remove the SMBH and its associated feedback in run $\texttt{NoBH}$.

\section{A self-regulated cool-core cluster}
\label{sec:selfreg}

\begin{figure*}
	\centering
	\includegraphics[trim=0cm 0cm 0cm 0cm,clip=true, width=0.48\textwidth]{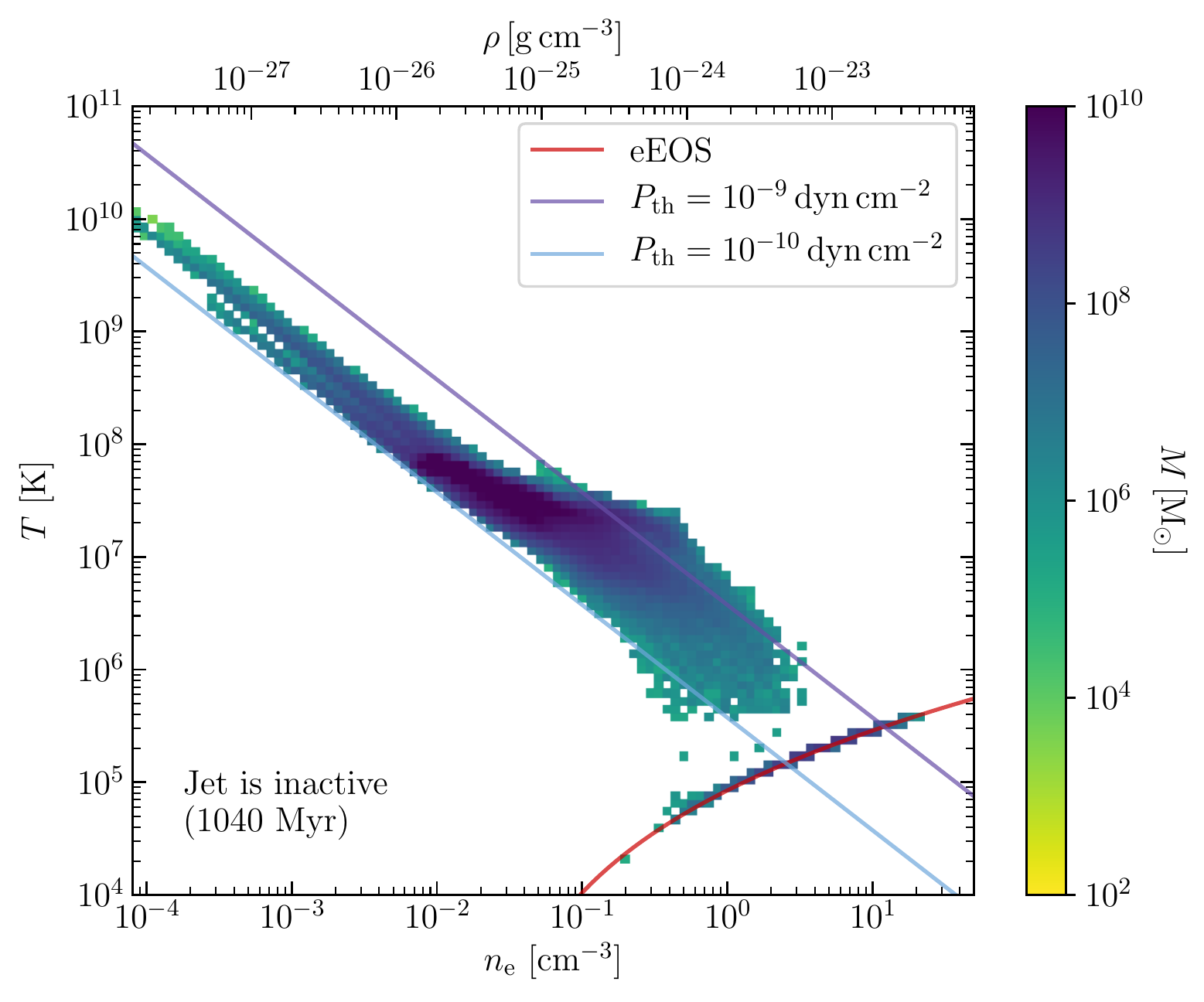}
	\includegraphics[trim=0cm 0cm 0cm 0cm,clip=true, width=0.48\textwidth]{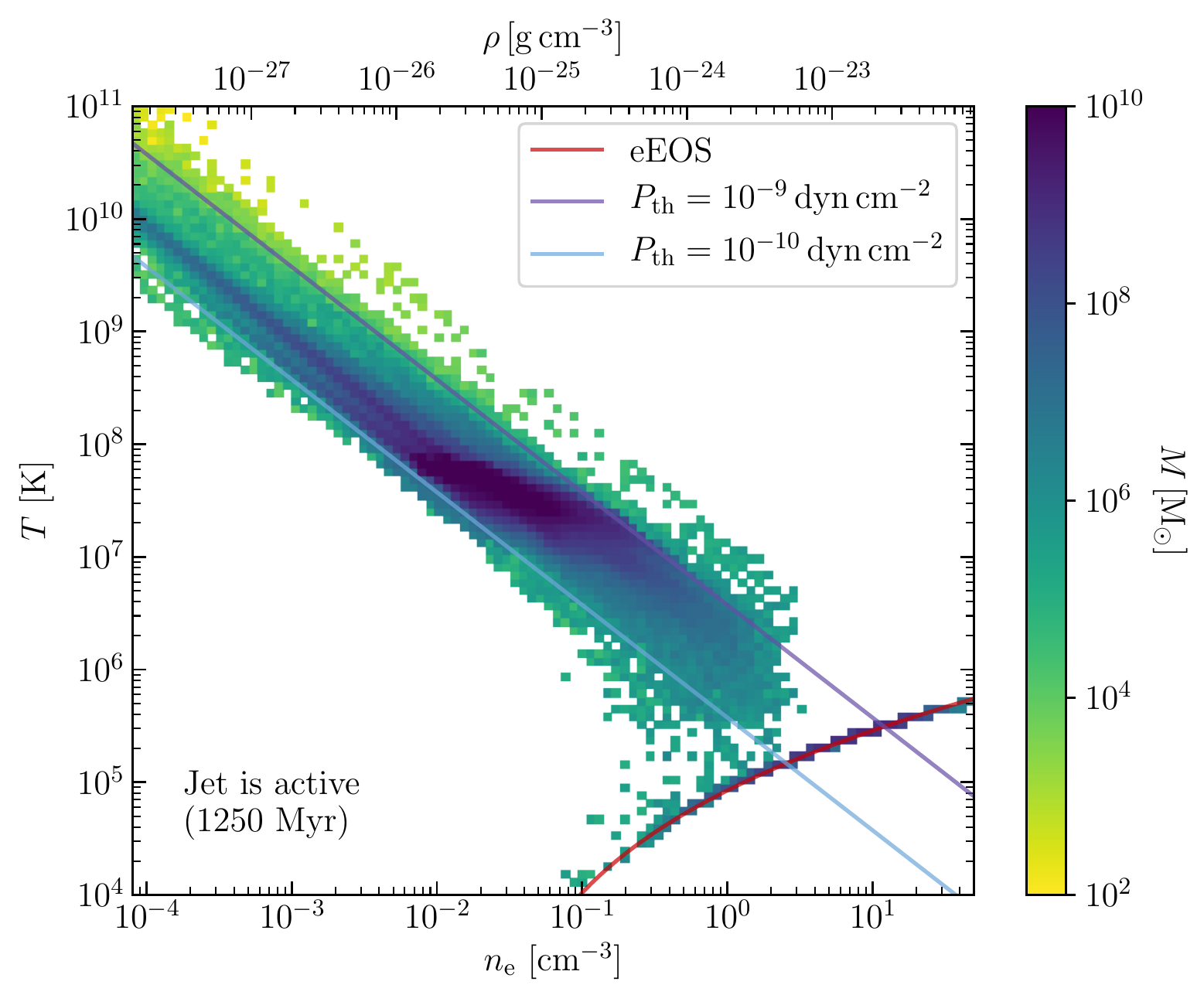}
	\caption{Phase diagram of the electron number density $n_\mathrm{e}$ and temperature in the central $100\,\mathrm{kpc}$ at $1040\,\mathrm{Myr}$ (where the jet is inactive) and at $1250\,\mathrm{Myr}$ (where the jet is active) in the run \texttt{Fiducial}. Color coding corresponds to total mass in respective bins. The effective model of the ISM inhibits cooling below the effective equation of state (red line). The plasma is near pressure equilibrium. Lines of constant pressure are shown in blue and purple.}
	\label{fig:phasediagram}
\end{figure*}

\begin{figure*}
	\centering
	\includegraphics[trim=0cm 0cm 0cm 0cm,clip=true, width=0.95\textwidth]{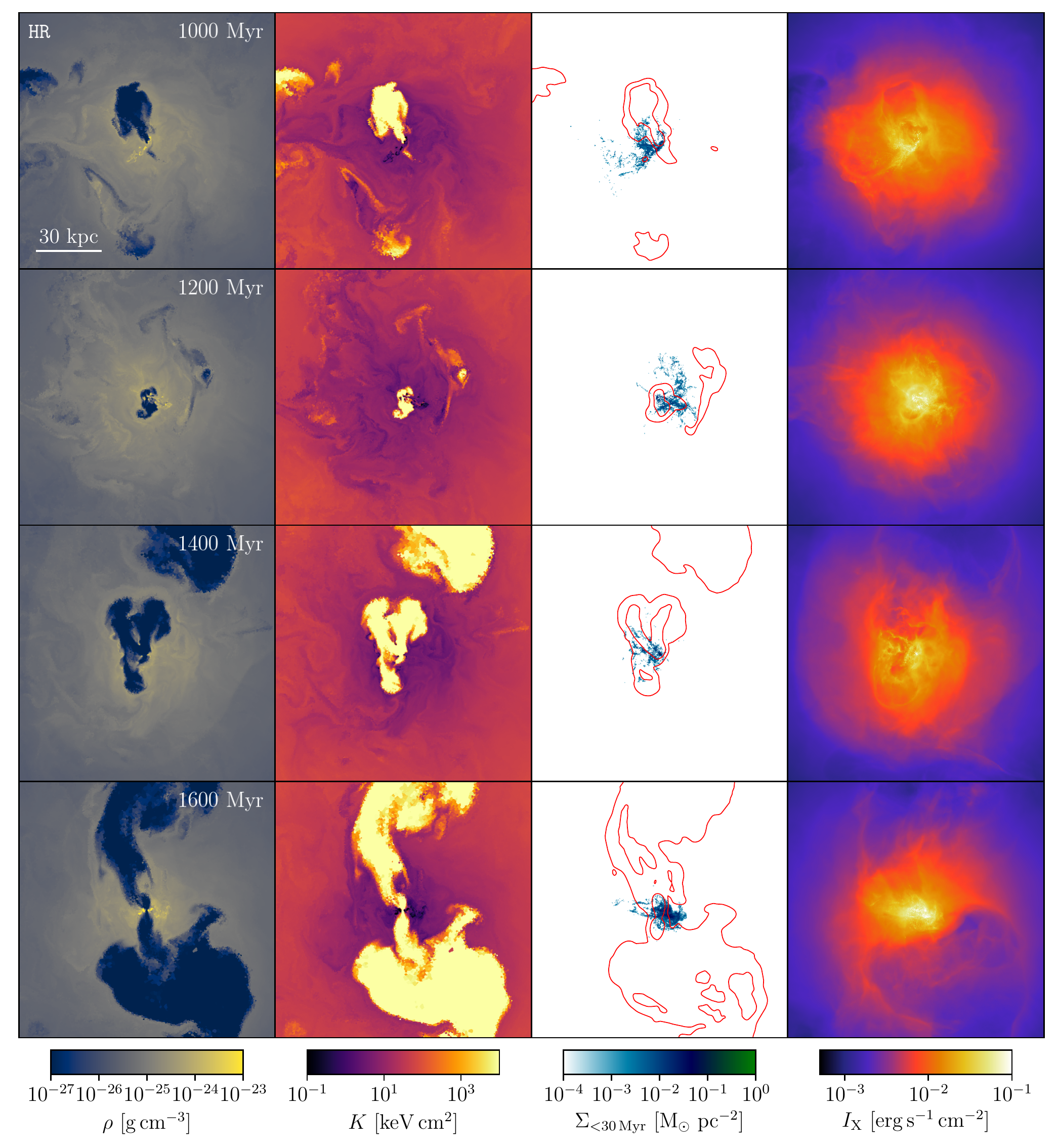}
	\caption{Overview of the fiducial high resolution cold accretion simulation \texttt{HR} (with accretion parameters $\epsilon=0.1$ and  $\eta=0.01$). We show thin slices of side length $120\,\mathrm{kpc}$ of the mass density $\rho$, entropy $K=k_\mathrm{B}Tn_\mathrm{e}^{-2/3}$, cold gas surface density $\Sigma_{<30\,\mathrm{Myr}}$ where $t_\mathrm{cool}<30\,\mathrm{Myr}$ and the X-ray emissivity $I_\mathrm{X}$ is in the Chandra band (for which we adopt our simulated cooling luminosity that includes metal line cooling). Cold gas deflects the jet which allows it to heat the cooling ICM more isotropically. High entropy gas is only found in the bubbles, whereas the bulk of the ICM remains at low entropy typical for CC clusters.}
	\label{fig:overviewhighres}
\end{figure*}

\begin{figure*}
	\centering
	\includegraphics[trim=0cm 0cm 0cm 0cm,clip=true, width=0.99\textwidth]{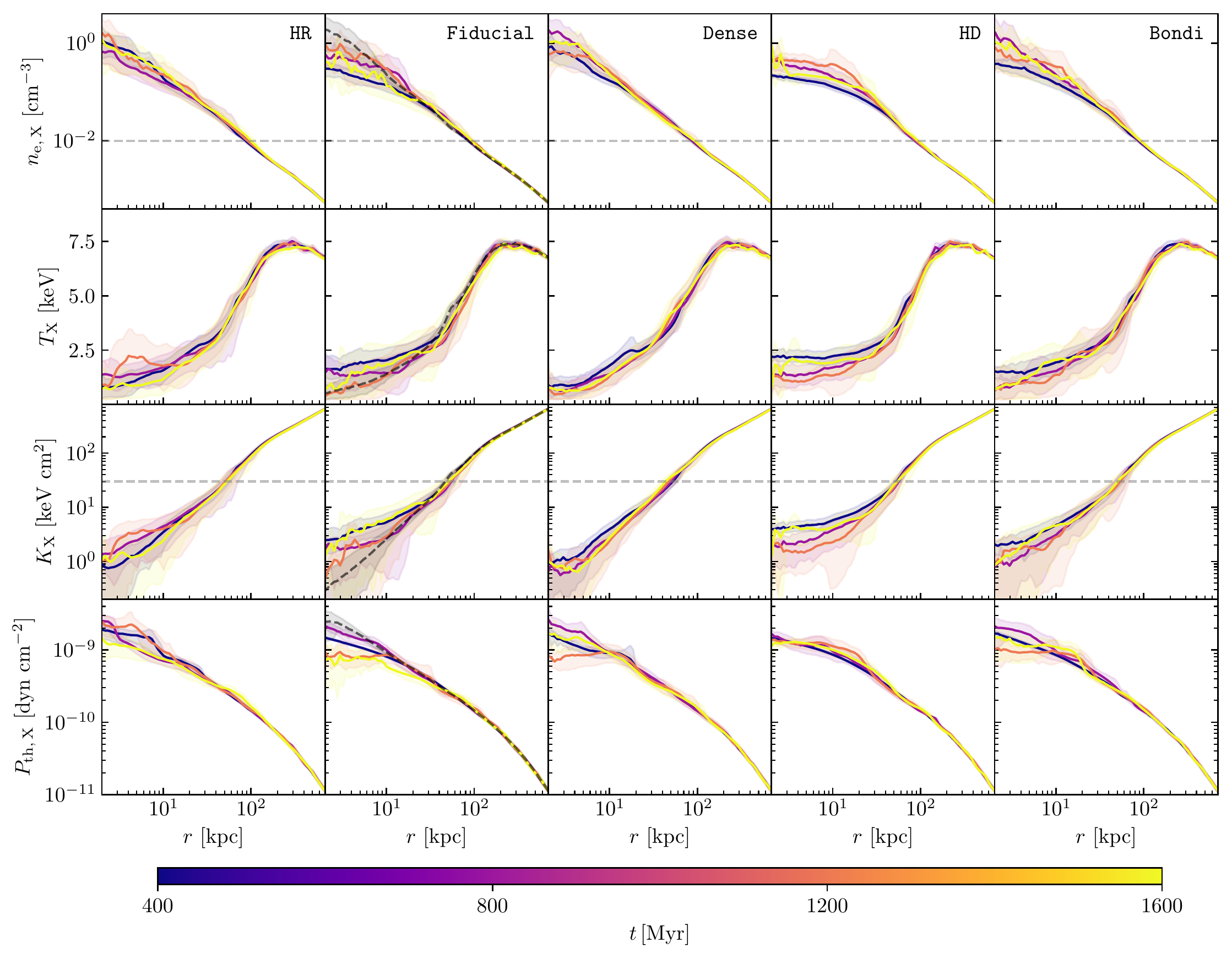}
	\caption{Radial profiles of the X-ray weighted density $n_{\mathrm{e},\,\mathrm{X}}$, temperature $T_{\mathrm{X}}$, entropy $K_{\mathrm{X}}$ and thermal pressure $P_{\mathrm{th},\,\mathrm{X}}$. The different colours correspond to simulation times as indicated on the colour bar. Profiles for model \texttt{NoBH} at $1200\,\mathrm{Myr}$ are shown in dashed in the second column. Shaded areas indicate the 10th to 90th percentiles and dashed lines represent popular choices for defining CC clusters (with central densities above $10^{-2}~\rmn{cm}^{-3}$ and central entropies less than $30~\rmn{keV cm}^2$). A self-regulated heating-cooling cycle leads to a dynamical attractor solution resembling that of observed CC clusters irrespective of the accretion mode and for all probed parameters.}
	\label{fig:radialprofilescc}
\end{figure*}

\begin{figure*}
	\centering
	\includegraphics[trim=0cm 0cm 0cm 0cm,clip=true, width=0.99\textwidth]{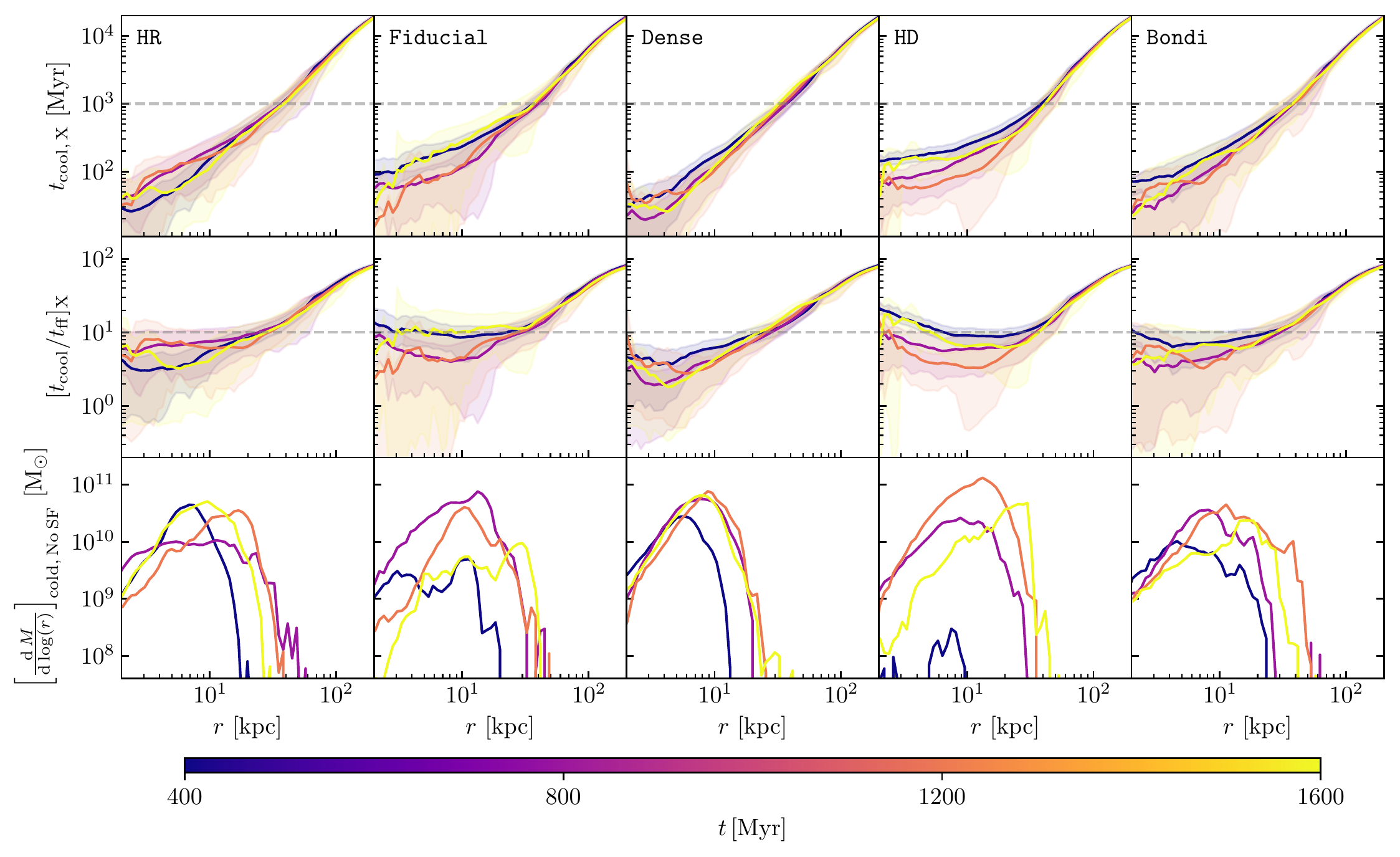}
	\caption{Radial profiles show the X-ray weighted cooling time $t_\mathrm{cool}$, the ratio of cooling-to-free fall time $t_\mathrm{cool}/t_\mathrm{ff}$ and the total mass per bin normalized by the bin size for non-star forming gas with $\tcool<100$~Myr, i.e.,  $\left[{\mathrm{d}M}/{\mathrm{d}\,\mathrm{log}(r)}\right]_{\mathrm{cold},\mathrm{No}\,\mathrm{SF}}$. The different colours correspond to simulation times as indicated on the colour bar. Shaded areas indicate the 10th to 90th percentiles. The ratio $t_\mathrm{cool}/t_\mathrm{ff}$ falls below 10 in the central region ($r\lesssim50\,\mathrm{kpc}$), where cold gas forms stars. The minimum in $t_\mathrm{cool}/t_\mathrm{ff}$ roughly corresponds to the maximum amount of cooling gas mass.}
	\label{fig:radialprofilescooling}
\end{figure*}

\begin{figure*}
	\centering
	\includegraphics[trim=0cm 0cm 0cm 0cm,clip=true, width=0.79\textwidth]{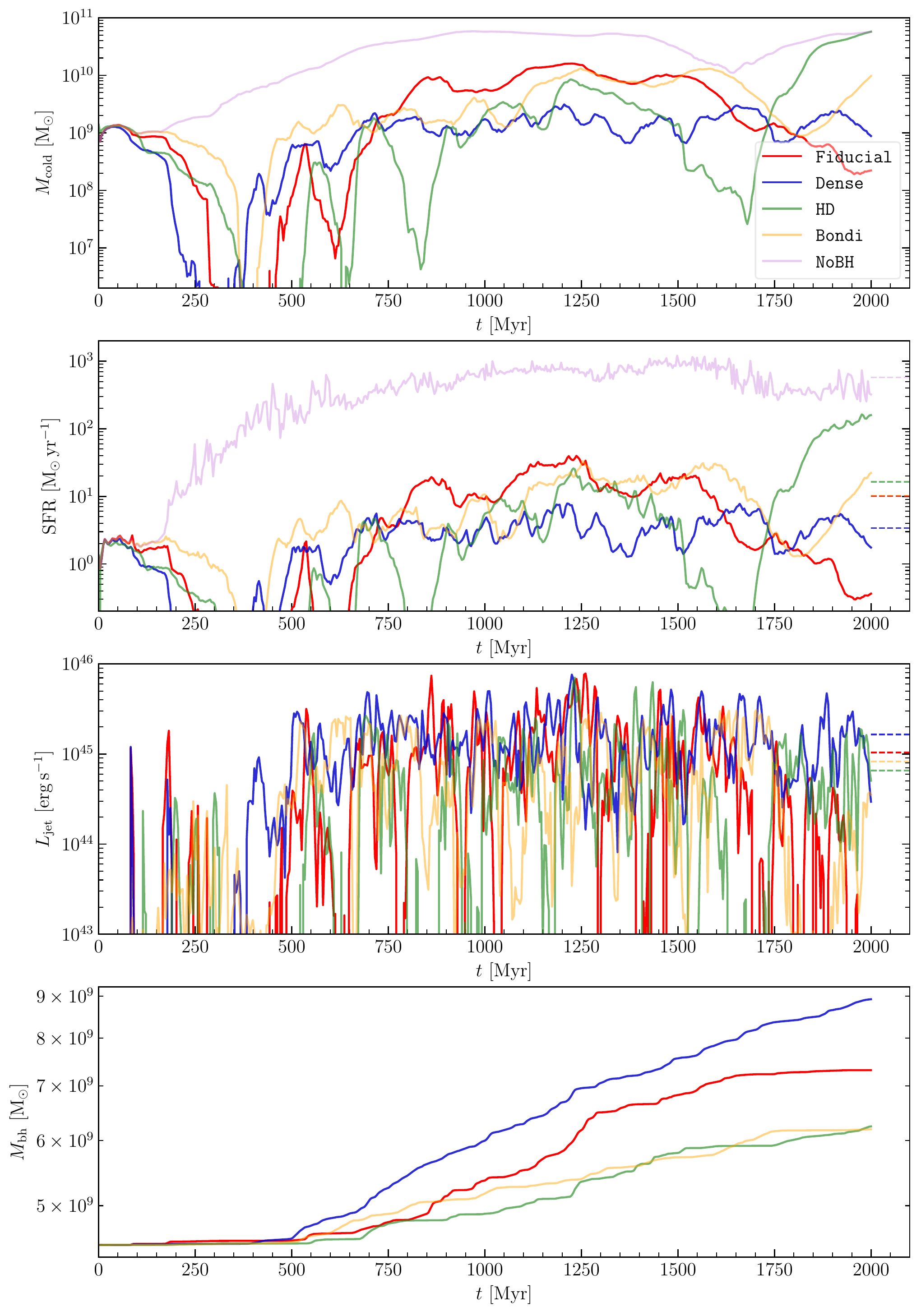}
	\caption{Time evolution of cold gas mass $M_\mathrm{cold}$ ($T<10^6\,\mathrm{K}$), star formation rate $\mathrm{SFR}$, jet power $L_\mathrm{jet}$ and SMBH mass $M_\mathrm{bh}$. Averages for times $t>500\,\mathrm{Myr}$ are shown with dashed horizontal lines on the right. If no feedback is included (\texttt{NoBH}) runaway cooling is observed and unobserved star formation rates of $>100\,\mathrm{M}_\odot\,\mathrm{yr}^{-1}$. After 500 Myr all feedback models establish a self-regulated state with moderate star formation rates ($\sim10\,\mathrm{M}_\odot\,\mathrm{yr}^{-1}$) and jet powers ($\sim8\times10^{44}\,\mathrm{erg}\,\mathrm{s}$). Dense jets have a larger momentum density and deposit their energy at larger radii so that the central gas cools more strongly and gives rise to larger SMBH masses in comparison to the other models. The hydrodynamical model forms an unobserved type of massive disc at $2\,\mathrm{Gyr}$ with extreme star formation rates ($\sim100\,\mathrm{M}_\odot\,\mathrm{yr}^{-1}$). In Fig.~\ref{fig:jetpowerhistogram} we show histograms of the jet power.}
	\label{fig:wangp}
\end{figure*}

\begin{figure*}
	\centering
	\includegraphics[trim=0cm 0cm 0cm 0cm,clip=true, width=1.000\textwidth]{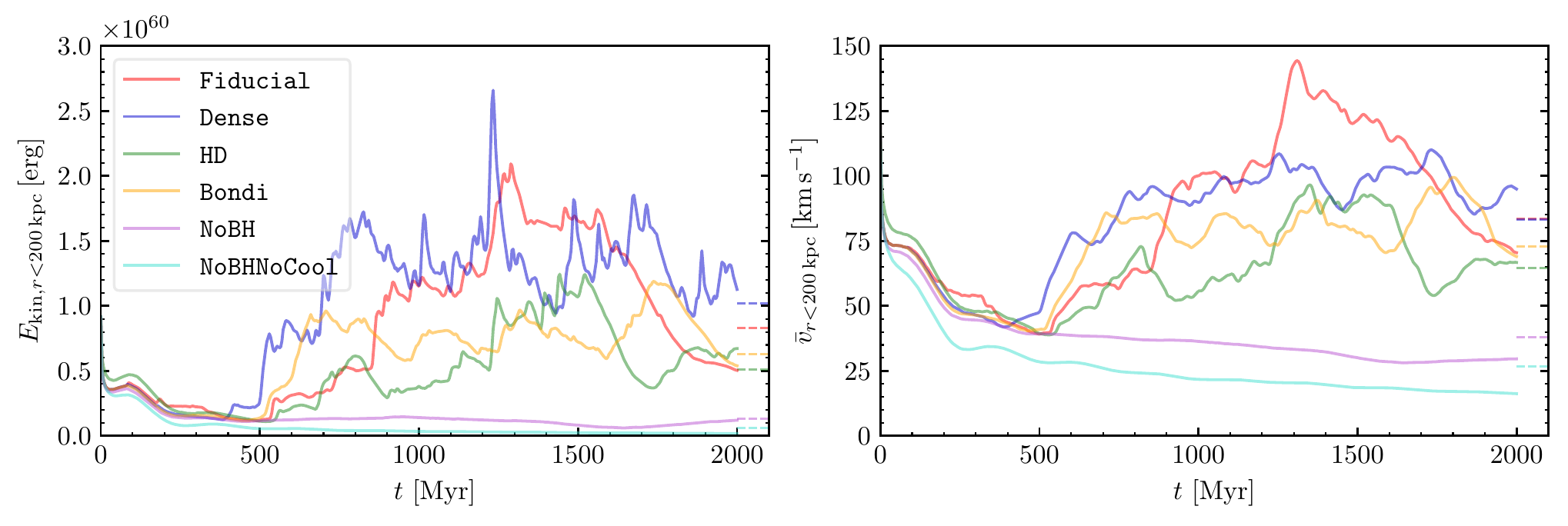}
	\caption{Time evolution of the total kinetic energy within $200\,\mathrm{kpc}$ ($E_{\mathrm{kin},r<200\,\mathrm{kpc}}$) and mass-weighted average velocity within $200\,\mathrm{kpc}$ ($\bar{v}_{r<200\,\mathrm{kpc}}$). Averages for all times are shown with a dashed horizontal lines on the right. Turbulent velocities that are present in the initial conditions decay as a function of time. The onset of jet feedback at $t>500\,\mathrm{kpc}$ in runs \texttt{Fiducial}, \texttt{Dense}, \texttt{HD} and \texttt{Bondi} injects substantial kinetic energy. On the other hand, runs excluding AGN feedback \texttt{NoBH} and \texttt{NoBHNoCool} (with and without radiative cooling, respectively) continue to lose kinetic energy. }
	\label{fig:Ekin_evolution}
\end{figure*}

\begin{figure*}
	\centering
	\includegraphics[trim=0cm 0cm 0cm 0cm,clip=true, width=0.98\textwidth]{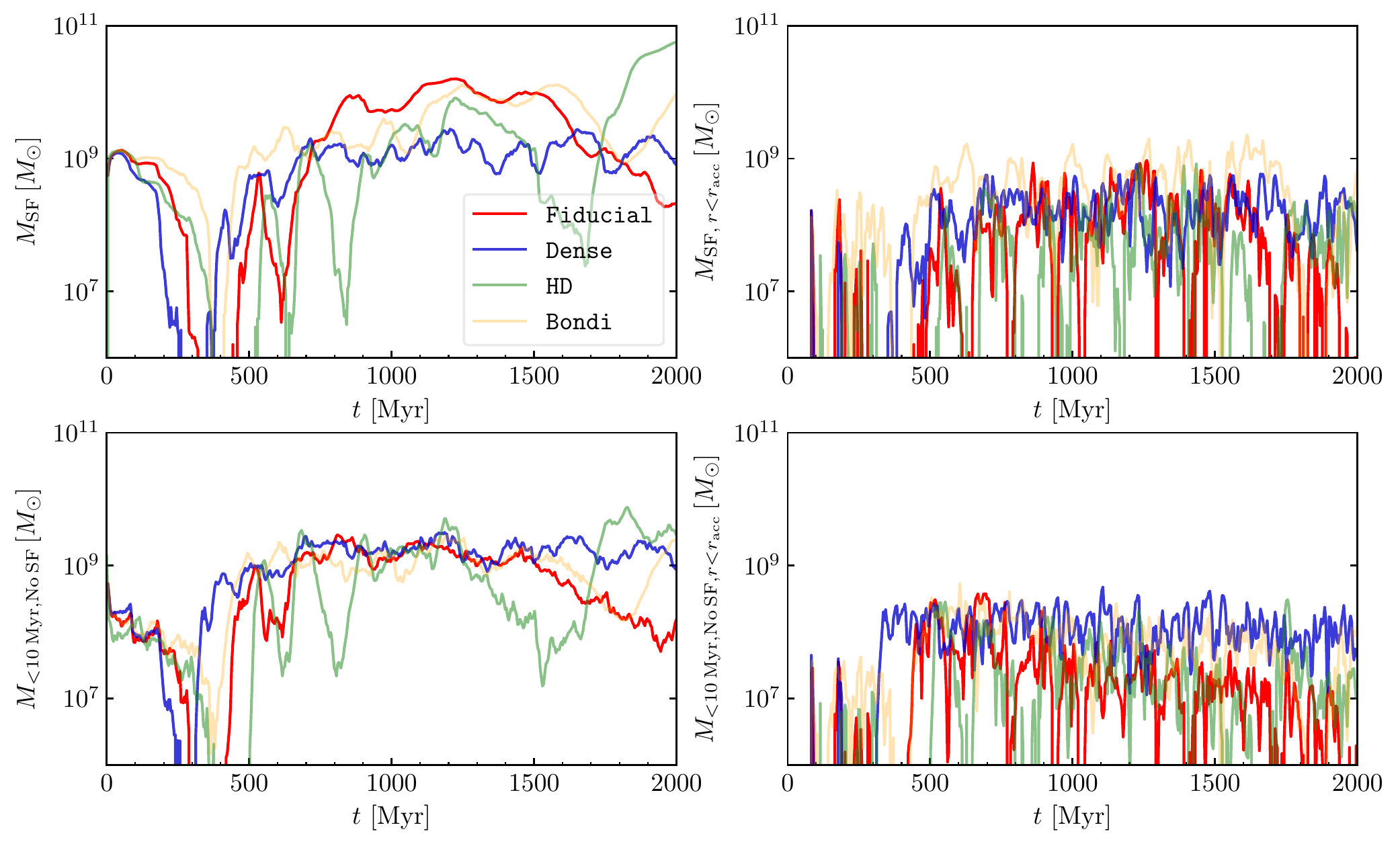}
	\caption{Mass in star forming gas (top) and gas with $t_\mathrm{cool}<10\,\mathrm{Myr}$ (bottom) within the entire simulation (left) and inside the accretion radius, $r<r_\mathrm{acc}$ (right). While cold and star forming gas is present throughout the simulation, its presence in the central accretion region is intermittent and varies with time. The gas is dragged up by the AGN and accelerated away from the center.}
	\label{fig:gasphases}
\end{figure*}

We first focus on the state of self-regulation seen in our simulations. The time evolutions of the different models are exemplified in Fig.~\ref{fig:overviewhighres}, where we show density and entropy slices, the surface density of the gas with a cooling time below $30\,\mathrm{Myr}$ and the X-ray emissivity of gas with $0.2\,\mathrm{keV}<k_\rmn{B}T<10\,\mathrm{keV}$ of our \texttt{HR} model ($\epsilon=0.1$, $\eta=0.01$, cold accretion). Low density jets inflate cavities with density contrasts of $\sim10^4$ that rise buoyantly in the cluster atmosphere.\footnote{While the central electron number density in the initial conditions is $5\times 10^{-2}~\rmn{cm}^{-3}$ (Eq.~\ref{eq:ne}), successive cooling and compression increases the central electron density so that it self-regulates around a new equilibrium profile with central densities of $\sim0.3$--$1~\rmn{cm}^{-3}$ (see Fig.~\ref{fig:radialprofilescc}), which implies ICM-to-jet density contrasts of $\sim3\times10^3$--$10^4$ in our fiducial low-density jets (and $\sim10$ in our \texttt{Dense} model) at jet launching.} Cluster weather resulting from turbulent intracluster motions deflects bubbles throughout the run. Additionally, central cold gas deflects the forming bubbles early which leads to more isotropised feedback. Deflection of bubbles by cluster weather and cold gas is seen in many simulations \citep[e.g.,][]{Sijacki2008,Morsony2010,Mendygral2012,Bourne2017,Bourne2019}. The gas in the wake of bubbles experiences a strong uplift, which advects low-entropy gas from the center to larger radii so that the upwards path leading to the high-entropy bubbles is traced by low-entropy gas that originates from the center \citep[as discussed in][]{chen2019,Ehlert2021,Zhang2022}. Gas with low cooling time accumulates in the center and forms filamentary structures and/or transient discs (e.g.\ at $1600\,\mathrm{Myr}$). The bubbles are clearly discernible as high-contrast cavities in the X-ray emissivity.

In Fig.~\ref{fig:phasediagram}, we show a phase diagram of temperature $T$ vs. electron number density $n_\mathrm{e}$ when the jet is inactive (left) and active (right) in the central $100\,\mathrm{kpc}$ of the cluster. As our low-density jets are set up in pressure equilibrium with the surrounding ICM, the jets are comprised of hot, low-density gas. Interestingly, the ICM pressure in the cluster center only varies by roughly an order of magnitude. This implies that jet feedback does not create any dramatically over-pressured gas in galaxy clusters, which, in turn, would cause strong shocks \citep[see the parameter study in][]{Ehlert2018}. Cooling gas moves isobarically onto the effective equation of state (eEOS) where it forms stars. Further cooling to higher densities and lower temperatures is limited by the eEOS.

We study the long-term impact of AGN feedback on the ICM in Fig.~\ref{fig:radialprofilescc}, where we show radial profiles of the X-ray weighted ($0.2\,\mathrm{keV}<k_\rmn{B}T<10\,\mathrm{keV}$) density $n_{\mathrm{e},\,\mathrm{X}}$, temperature $T_{\mathrm{X}}$, entropy $K_{\mathrm{X}}$ and thermal pressure $P_{\mathrm{th},\,\mathrm{X}}$ of various runs. Dashed lines separate strong CC clusters from moderate CC/non CC clusters as observationally determined \citep{Cavagnolo2008,Hudson2010}. Central densities remain high throughout time independent of accretion model, resolution and jet density, well above the CC limit. The initialization of denser jets in our \texttt{Dense} model lead to even higher densities in this run. Generally, the central entropy stays well below $10\,\mathrm{keV}\,\mathrm{cm}^2$. As noted previously, radiative cooling of our initial conditions causes the central temperature to drop by a factor of $\sim3$, the density to increase by a factor of a few and therefore the entropy to decrease by a factor $\sim10$.

Interestingly, the cluster self-regulates at these new equilibrium profiles throughout the runtime of the simulation, implying that our AGN jet feedback stabilises the system but cannot substantially alter the thermodynamic profile of our CC galaxy cluster or even transform it into a non-CC cluster \citep[in agreement with the statistics of observed AGN bubble enthalpies and central ICM entropies,][]{Pfrommer2012}. For comparison, we show profiles for model \texttt{NoBH} at $1200\,\mathrm{Myr}$ in the second column of Fig.~\ref{fig:radialprofilescc}, which enters a run-away cooling state with a dense cold core that causes SFRs elevated by nearly two orders of magnitude over the other models. More work with different initial conditions and cosmological settings is needed to test the universality of this prediction. Increasing the initial density and introducing a more granulated density structure as seen in turbulent box simulations \citep[e.g.,][]{Mohapatra2021a} may lead to equilibrium densities at the observed levels \citep[see Eq.~\eqref{eq:ne},][]{Churazov2003}.

To explore the cooling gas, we show in Fig.~\ref{fig:radialprofilescooling} radial profiles of X-ray weighted ($0.2\,\mathrm{keV}<k_\rmn{B}T<10\,\mathrm{keV}$) cooling time $t_{\mathrm{cool,X}}$ and the cooling-to-free fall time ratio $\left[t_\mathrm{cool}/t_\mathrm{ff}\right]_\mathrm{X}$. In addition, we plot the total mass per logarithmic bin, $\left[{\mathrm{d}M}/{\mathrm{d}\,\mathrm{log}(r)}\right]_{<100\,\mathrm{Myr},\mathrm{No}\,\mathrm{SF}}$, for non-star-forming gas with $\tcool<100$~Myr. The cooling times stay below $1\,\mathrm{Gyr}$ within the inner $50\,\mathrm{kpc}$ independent of the accretion model, resolution and jet density. The cooling-to-free fall time ratio remains low within $r\lesssim50\,\mathrm{kpc}$ at $\left[t_\mathrm{cool}/t_\mathrm{ff}\right]_\mathrm{X}<10$. For $t>400\,\mathrm{Myr}$ cold gas is present in the center at $r\lesssim30\,\mathrm{kpc}$. This state of cooling appears to be a general feature of our simulations.

In Fig.~\ref{fig:wangp}, we show the time evolution of the cold gas mass ($M_\mathrm{cold}$ with $T<10^6\,\mathrm{K}$), the star formation rate ($\mathrm{SFR}$), the jet luminosity ($L_\mathrm{jet}$) and the SMBH mass ($M_\mathrm{bh}$).
Due to the Gaussian temperature fluctuations in the initial conditions, cold gas collapses in multiple clumps at $r\sim20\,\mathrm{kpc}$. The clumps fall towards the center but due to their angular momentum, they overshoot in their orbits. Initial SMBH accretion is thus limited to short bursts when the clumps pass near the SMBH.
After $\sim500\,\mathrm{Myr}$, our AGN jet feedback establishes a cycle of self-regulation on characteristic timescales throughout individual runs with similar jet powers and SFRs for each run. Therefore, we focus our analysis on times after $500\,\mathrm{Myr}$. SFRs are closely tied to cold gas formation. Runs including AGN jet feedback show $\mathrm{SFR}\sim10\,\mathrm{M}_\odot\,\mathrm{yr}^{-1}$ while our \texttt{NoBH} model reaches $\mathrm{SFR}>400\,\mathrm{M}_\odot\,\mathrm{yr}^{-1}$. The \texttt{Dense} model shows systematically larger SMBH accretion rates. The increased accretion rates imply larger SMBH masses and AGN jet luminosities (by a factor of $\sim2$, see dashed lines in Fig.~\ref{fig:wangp}) so that the SFR is more quenched in comparison to the other models. As we will show below, this model enables the formation of long-lived discs that continuously feed the SMBH.

For runs including jet feedback, the cold gas masses mostly stay in the range $10^9\,\mathrm{M}_\odot<M_\mathrm{cold}<10^{10}\,\mathrm{M}_\odot$ after $t>500\,\mathrm{yr}$. Only the model \texttt{NoBH} forms significantly more cold gas, which implies a substantially increased SFR. Towards the end of our  \texttt{HD} simulation, we observe a large amount of cold gas, which results from a forming massive gas disc with radius $r>5\,\mathrm{kpc}$ in the center. Cold gas is trapped on circular orbits and cannot be accreted by the SMBH to fuel the feedback cycle. \cite{Li2014b} observe a similar disc in their hydrodynamical simulations.

After time $t~>~500\,\mathrm{Myr}$, the jet power hovers between $10^{44}\,\mathrm{erg}\,\mathrm{s}^{-1}<L_\mathrm{jet}<10^{46}\,\mathrm{erg}\,\mathrm{s}^{-1}$, in agreement with observational power estimates that are required to inflate X-ray cavities in a sample of CC clusters \citep[e.g.,][]{Rafferty2006}. While jets in \texttt{Bondi} and \texttt{Dense} remain active throughout the simulation time, \texttt{Fiducial} and \texttt{HD} show more intermittent jet behavior. The jet luminosities in the \texttt{Dense} model are on average twice as high as in the \texttt{Fiducial} model. In the \texttt{Fiducial} model, the low density jets are easily deflected by cold central clumps, yielding drastically varying jet directions. The jets with a much higher momentum density in the \texttt{Dense} model, however, keep their direction. Final SMBH masses reach $\mathrm{M}_\mathrm{bh}\sim7-9\times10^9\,\mathrm{M}_\odot$. Consequently, SMBHs grow at most by a factor of 3 over $2\,\mathrm{Gyr}$ in line with expectations.

Figure~\ref{fig:Ekin_evolution} shows the time evolution of the total kinetic energy and mass-weighted average velocity within $200\,\mathrm{kpc}$. The initial velocity fluctuations quickly decay as our non-cosmological set-up does not support continuous driving as a result of gravitational potential fluctuations owing to gas accretion and mergers. Once AGN feedback sets in (at $t>500\,\mathrm{kpc}$), kinetic energy is injected into the central regions that self-regulates at a level of $E_{\mathrm{kin},r<200\,\mathrm{kpc}}\sim(0.5$--$1)\times10^{60}$~erg. This generates mass-weighted average velocities at the level of around $\bar{v}_{r<200\,\mathrm{kpc}}\approx 75\,\rmn{km~s}^{-1}$, with fluctuations of a factor of two, which are qualitatively similar in our different AGN feedback models (\texttt{Fiducial}, \texttt{Dense}, \texttt{HD} and \texttt{Bondi}). Clearly, simulations without AGN feedback (\texttt{NoBH} and \texttt{NoBHNoCool}) continue to lose kinetic energy. However, radiative cooling leads to a drop in thermal pressure, which initiates motions in the ICM (model \texttt{NoBH}), which partially compensates for the turbulent decay of kinetic energy and thus sustains larger velocities in comparison to our non-radiative model (\texttt{NoBHNoCool}).

In Fig.~\ref{fig:gasphases}, we show the total mass in star forming gas (top) and non-star-forming gas with $t_\mathrm{cool}<10\,\mathrm{Myr}$ (bottom) while we compare the gas reservoirs in the entire simulation (panels on the left-hand side) and within the accretion radius (panels on the right-hand side). Whereas the amount of star formation and cold gas varies with time, the cluster remains in a state of constant cooling and star formation. However, if the presence of cold gas is intermittent in the accretion region, the SMBH accretion rate reflects this behaviour. Especially in the runs \texttt{HD} and \texttt{Fiducial}, cold gas and star forming gas are temporarily absent in the accretion region.

\section{Connecting jet activity to observations of X-ray cavities}
\label{sec:cavities}
Jet activity on tens of Myr timescales in CC clusters can be constrained by measuring the energy contained in hot bubbles inflated by AGN driven jets and observed as X-ray cavities. Here, we want to compare the cavity luminosity with the instantaneous jet power and cooling luminosity obtained from the simulations.

\begin{figure*}
	\centering
	\includegraphics[trim=0cm 0cm 0cm 0cm,clip=true, width=0.8\textwidth]{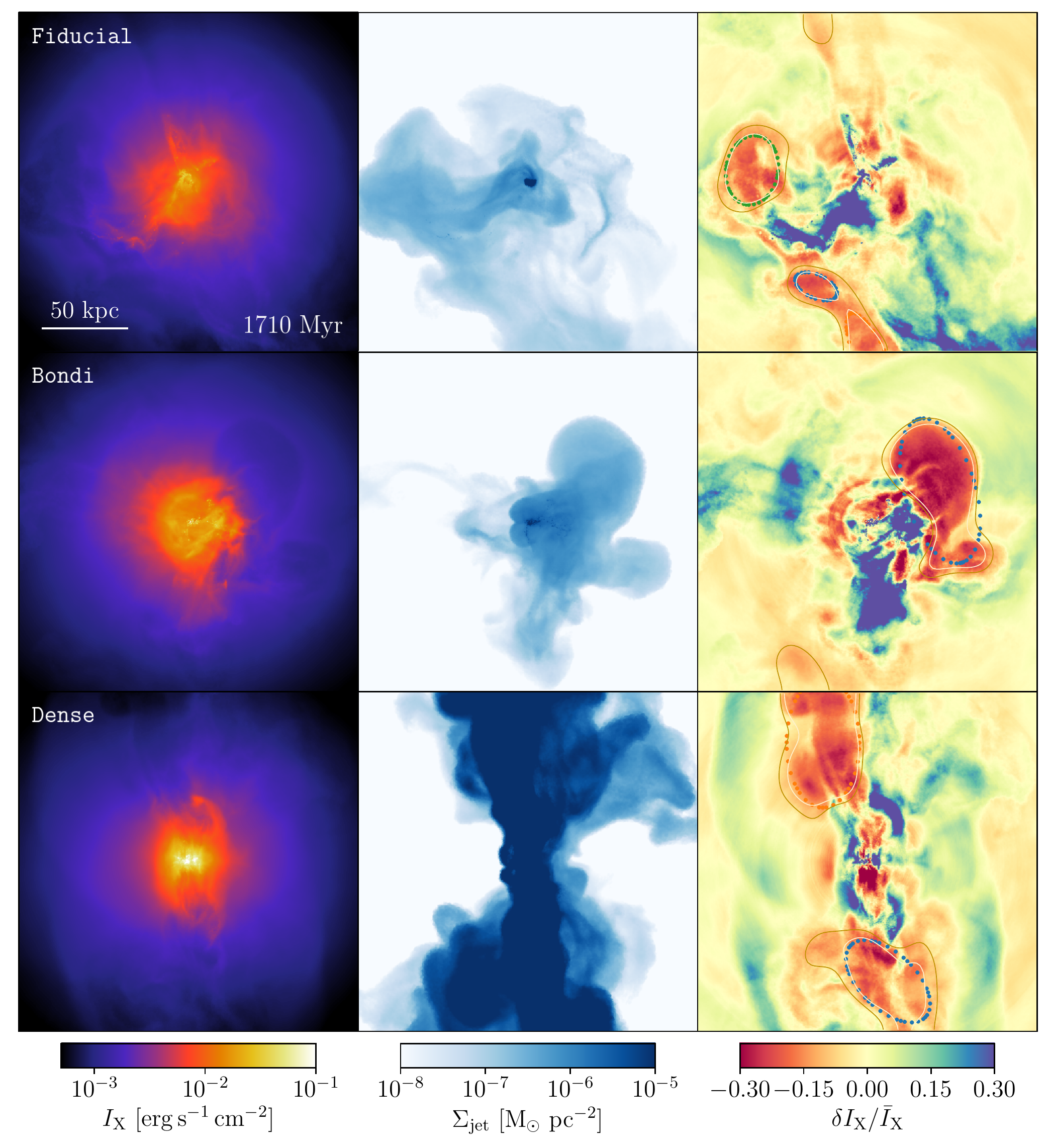}
	\caption{X-ray emissivity in the Chandra band (left column), integrated jet mass surface density (central column), and $\delta I_\rmn{X}/\bar{I}_\rmn{X}$ (right column, at a resolution of $512\,\mathrm{pixels}\times512\,\mathrm{pixels}$). The projections maps have been integrated through the entire simulation box, the size of the images are $200\,\mathrm{kpc}$, and we show a simulation time of $1560\,\mathrm{Myr}$. We devise an automated algorithm for finding cavities in the X-ray images by (i) subtracting an azimuthally symmetric profile $I_\rmn{s}$ of the X-ray emissivity and normalizing it with the mean profile, $\delta I_\rmn{X}/\bar{I}_\rmn{X}$, and (ii) fitting ellipses to cavities with a contrast of 15 per cent in the residual maps. Images used for this methods have a resolution of $100\,\mathrm{pixels}\times100\,\mathrm{pixels}$ and a shallower depth (see Section.~\ref{sec:fittingxraycavities} for details). Cavities in the resulting image at $10$ per cent (red) and $15$ per cent (gray) are marked with contours and the fitted ellipses are shown with dots. The ellipse volume (where depth is assumed equal to the minor axis) and average pressure at its center are used to determine the cavity power $L_\mathrm{cav}$ (see Fig.~\ref{fig:jetpowerhistogram}).}
	\label{fig:cavityfittingexplanation}
\end{figure*}

\subsection{Fitting X-rays cavities in the simulations}
\label{sec:fittingxraycavities}
To efficiently identify cavities in our simulations with similar constraints as imposed by observations, we first compute the X-ray emissivity $I_\mathrm{X}$ within a SMBH-centered image with dimension $150\,\mathrm{kpc}\times150\,\mathrm{kpc}$ and depth $150\,\mathrm{kpc}$ (panels on the left-hand side in Fig.~\ref{fig:cavityfittingexplanation}). Defining the azimuthally averaged X-ray profile $\bar{I}_\rmn{X}$, we construct the X-ray fluctuation image via $(I_\mathrm{X}-\bar{I}_\rmn{X})/\bar{I}_\rmn{X}$. To simplify our fitting procedure, we use a coarse grained X-ray image with a bin size of $10\,\mathrm{kpc}$ and fit ellipses to contour lines with $\delta I_\rmn{X}/\bar{I}_\rmn{X}=-0.15$. This enables us to compute cavity energies $E_\mathrm{cav}=P_\rmn{th}V$, where the thermal pressure $P_\rmn{th}$ is the average value at the radius of the bubble center and $V$ is the ellipsoid volume by assuming the depth of the ellipsoid to be equal to the minor axis of the fitted ellipse. We reject any fitted cavities with centers outside a cube of side length of $150\,\mathrm{kpc}$. The cavity luminosity $L_\mathrm{cav}$ is then given by
\begin{equation}
	L_\mathrm{cav}=\frac{E_\mathrm{cav}}{t_\mathrm{rise}}=\frac{P_\rmn{th}V}{t_\mathrm{rise}},
\end{equation}
 where $t_\mathrm{rise}$ is the bubble rise time, which we assume to be equal to the sound crossing time $t_\mathrm{sc}$;
\begin{equation}
t_\mathrm{rise}=t_\mathrm{sc}=R\sqrt{\frac{\mu m_\mathrm{p}}{\gamma_\mathrm{a} k_\mathrm{B}T}}\approx40\,\mathrm{Myr}\left(\frac{R}{40\,\mathrm{kpc}}\right)\left(\frac{k_\mathrm{B}T}{3\,\mathrm{keV}}\right)^{-1/2},
\end{equation}
where $\gamma_\mathrm{a}=5/3$ for the ambient ICM.

The method is able to recover the most relevant cavities. Bubbles can disrupt into smaller cavities during their late-time evolution as a result of Kelvin-Helmholtz instabilities, which are suppressed by magnetic draping \citep{Dursi2007,Dursi2008,Ehlert2018}. An example of such a splitting of an AGN lobe into two smaller bubbles can be observed in the bottom-left corner of the top panel of Fig.~\ref{fig:cavityfittingexplanation}. Note that our algorithm with the detection threshold is tuned to bubbles inflated by our low-density (fiducial) jets. Therefore, denser bubbles inflated in our model \texttt{Dense} may therefore not be fully recovered, especially if they are small.

In the panels on the right-hand side in Fig.~\ref{fig:cavityfittingexplanation}, we show $\delta I_\rmn{X}/\bar{I}_\rmn{X}$ at a resolution of $512\,\mathrm{pixels}\times512\,\mathrm{pixels}$, projected through the entire simulation box. Low density regions and regions dominated by kinetic energy like the bubbles become discernible as a dip in relative intensity (red). These regions are well captured by our cavity finder. On the other hand, thermally compressed regions and shocks are visible as an increase in relative intensity (blue). Here, regions below the cavities show a strong excess in X-ray emissivity in the upper and central panels, while the lower panel shows bubbles in the process of being inflated, with signatures of shocks at the bubble-ICM interface. Those are currently not found with our cavity finder.

\subsection{Comparing cavity powers to jet powers and cooling luminosities}
\begin{figure*}
\centering
\includegraphics[trim=0cm 0cm 0cm 0cm,clip=true, width=0.99\textwidth]{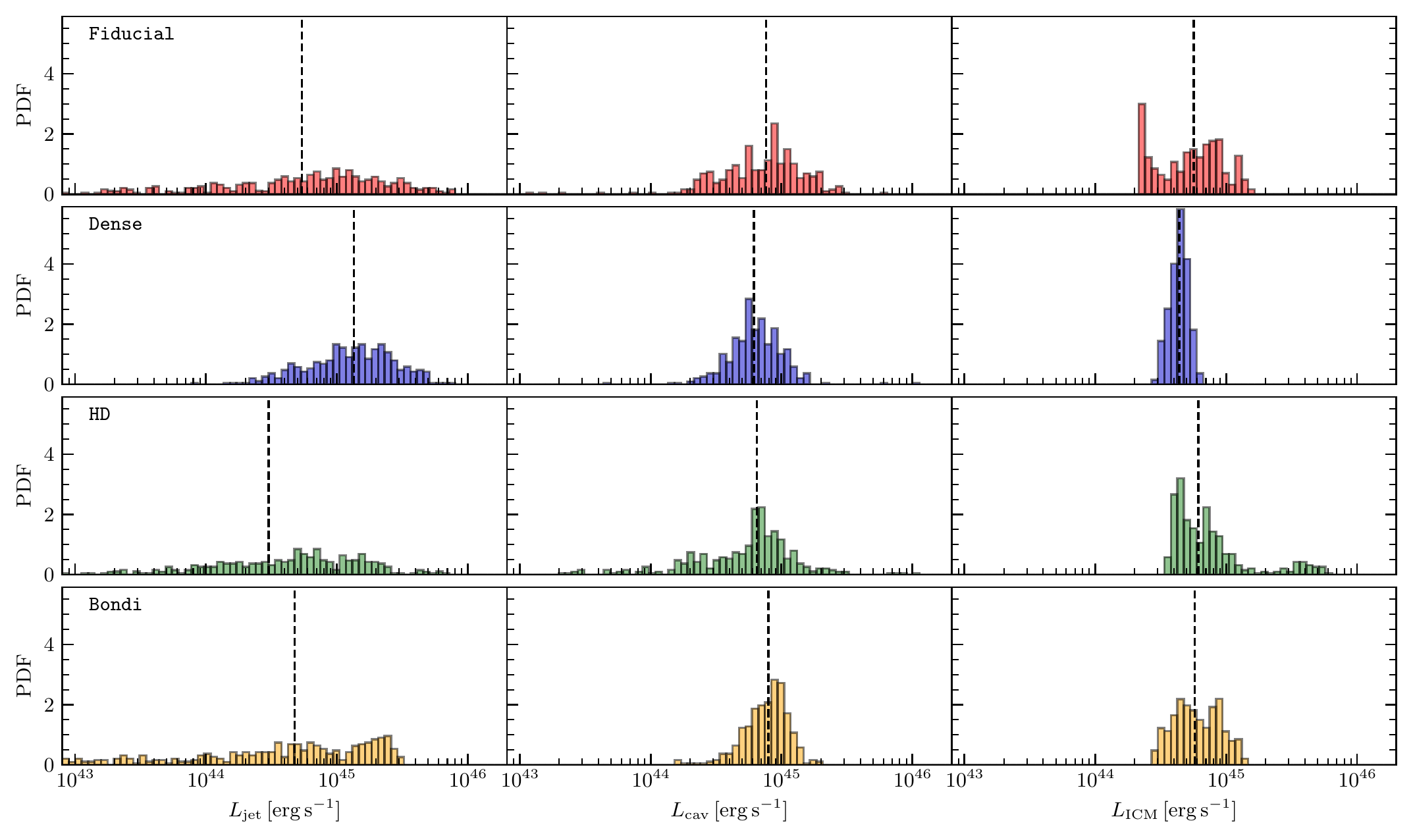}
\caption{Normalized histograms of the jet power, cavity power, and cooling power within 30 kpc (left to right) of the \texttt{Fiducial}, \texttt{Dense}, \texttt{HD}, and \texttt{Bondi} run after $500\,\mathrm{Myr}$. Dashed lines indicate median values. Low-power jet events ($L_\mathrm{jet}\lesssim10^{44}\,\mathrm{erg}\,\mathrm{s}^{-1}$) inflate cavities that are systematically missed by our cavity detection algorithm. While the distributions of cooling power and cavity power are very similar, we observe a broader distribution of jet powers with a tail towards low-luminosity events that are triggered by individual cooling filaments.}
    \label{fig:jetpowerhistogram}
\end{figure*}

Figure~\ref{fig:jetpowerhistogram} shows from left to right probability distribution functions (PDFs) of jet power, cavity powers and ICM cooling luminosities within the central $30\,\mathrm{kpc}$ for models indicated in the legends. Jet powers encompass a range between $10^{43}$--$10^{46}\,\mathrm{erg}\,\mathrm{s}^{-1}$, where the exact distribution is somewhat model dependent. The jet powers of model \texttt{Dense} show a lower scatter with an increased median (dashed lines) compared to the \texttt{Fiducial} model, which in turn shows a slightly higher median value than the models \texttt{HD} and \texttt{Bondi}. Interestingly, median cavity powers are much more similar at $\sim6$--$7\times10^{44}\,\mathrm{erg}\,\mathrm{s}^{-1}$ for all models. ICM luminosities also show a smaller scatter around median values of $\sim7$--$10\times10^{44}\,\mathrm{erg}\,\mathrm{s}^{-1}$.

\begin{figure*}
\centering
\includegraphics[trim=0cm 0cm 0cm 0cm,clip=true, width=0.99\textwidth]{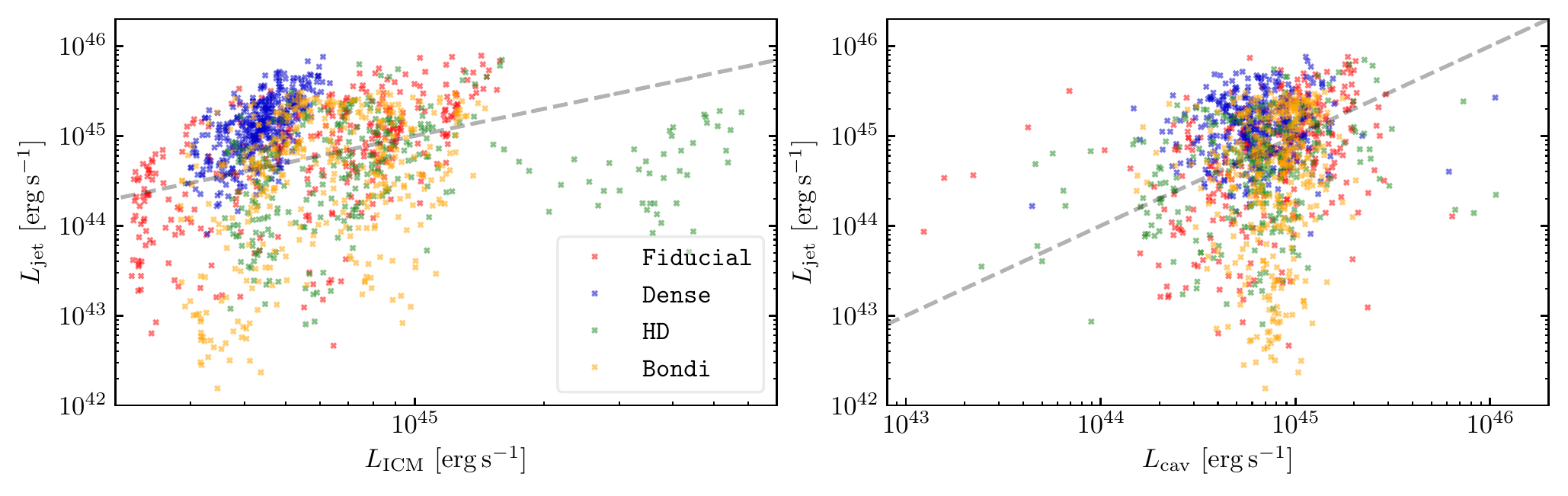}
\caption{Jet power $L_\mathrm{jet}$ vs.\ ICM cooling luminosity $L_\mathrm{ICM}$ (left) and jet power vs.\ cavity power (right) after $500\,\mathrm{Myr}$. Jet powers do not necessarily correspond to the current cooling luminosity. Self-regulation is achieved in an average sense on longer timescales. The hydrodynamical run shows an excursion of a phase of stronger cooling, especially at later times when a disc is formed. We note that our cavity powers span two orders in magnitude while the corresponding jet powers span four orders of magnitude.}
    \label{fig:cavityheatingcooling}
\end{figure*}

The main reason that these highly variable jet powers do not result in similar variances of the cavity powers is the high-frequency time variability of these powers. Because cavities are inflated over timescales of $\sim10\mathrm{s}\,\mathrm{Myr}$, the cavity powers correspond to average values over at least these timescales. Our algorithm is able to detect old cavities (with a start of the inflation $>100\,\mathrm{Myr}$) out to scales of $>50\,\mathrm{kpc}$, rising in the atmosphere. The continuous inflation of new bubbles with only short times of quiescence of $\sim50\,\mathrm{Myr}$ (cf.\ the evolution of $L_\mathrm{jet}$ in Fig.~\ref{fig:wangp}) causes bubbles to always exist in the cluster and to encompass a combined jet cavity power ($L_\mathrm{cav}$) that reaches similar values in our different models. In Fig.~\ref{fig:cavityheatingcooling}, we show jet power vs.\ cavity power on the right, which echoes these results. Jets with $L_\mathrm{jet}\lesssim10^{44}\,\mathrm{erg}\,\mathrm{s}^{-1}$ do not result in separate cavities but instead are contributing to inflating larger, higher luminous cavities. Note that our simulated cavity powers reflect observational estimates of cavity luminosities \citep{Birzan2004,Rafferty2006,Diehl2008}.

While the jet power evolutions in Figs.~\ref{fig:wangp} and \ref{fig:wangpmodelparameters} show periods of jet inactivity in some models, we find that cavities are always present throughout the runtime. This is consistent with observations of 55 cluster of which $60$--$100$ per cent show cavities \citep{Birzan2012}. Interestingly, we find that the jet powers $L_\mathrm{jet}$ exceed the central cooling luminosities $L_\mathrm{ICM}$ at most by an order of magnitude. However, jet powers that are up to 2 orders of magnitude lower than cooling luminosities are relatively common. Consequently, AGN feedback keeps the cluster in a state of moderate cooling with low but persistent star formation \citep[e.g.,][]{Voit2005a,Cavagnolo2009}.

\section{Magnetic coupling of cold and hot gas}
\label{sec:bfield_coldgas}
In this section, we study the morphology and kinematics of the cold gas and how magnetic fields influence it. We also address the relevance of magnetic fields in redistributing the AGN feedback energy and how it is related to the hot-phase observables, such as the velocity distribution of the X-ray emitting gas and the Faraday rotation measure.

\subsection{Impact of magnetic fields on cold gas morphology and kinematics}
\label{sec:coldgas}

\begin{figure*}
\centering
\includegraphics[trim=0cm 0cm 0cm 0cm,clip=true, width=0.95\textwidth]{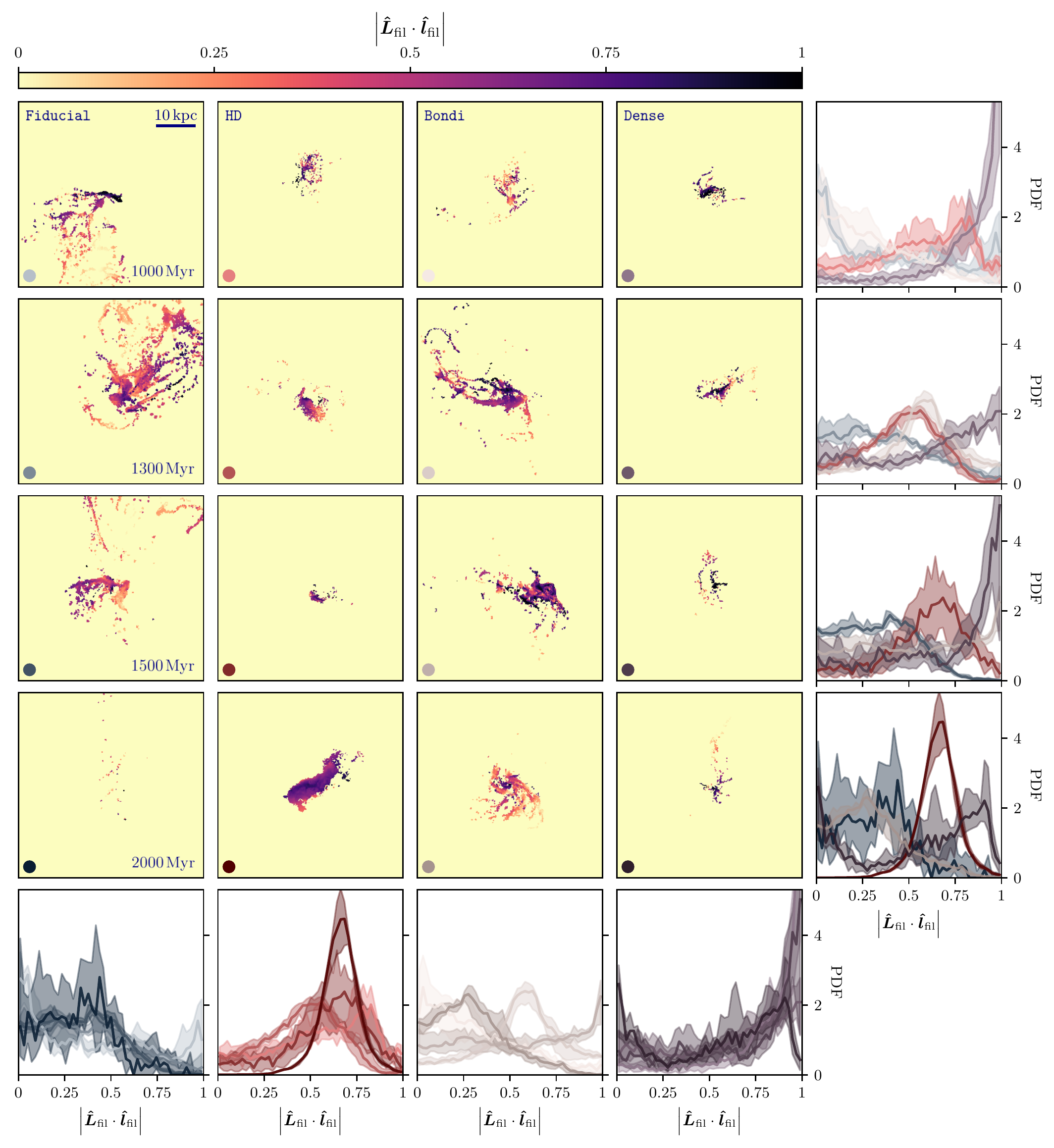}
\caption{Projection of normalized angle between total angular momentum of the cold gas and that of the individual cells $|\bm{\hat{L}}_\mathrm{fil}\bcdot\bm{\hat{l}}_\mathrm{fil}|$ at 1000, 1300, 1500 and 2000 Myr (top to bottom) for different runs indicated in the legends (left to right). The bottom (right-most) panels show the PDF of $|\bm{\hat{L}}_\mathrm{fil}\bcdot\bm{\hat{l}}_\mathrm{fil}|$ of the different panels in the respective columns (rows) with colors in the lower left corners of the panels identifying the corresponding PDFs. A discy distribution implies values of $|\bm{\hat{L}}_\mathrm{fil}\bcdot\bm{\hat{l}}_\mathrm{fil}|\approx 1$ while a random distribution is flat in this quantity.   We see a large variety of gas kinematics over time for the different models while the \texttt{HD} and -- to some extent -- the \texttt{Dense} models are significantly discier than our MHD models with light jets. }
\label{fig:epsilonplot}
\end{figure*}

The dynamics in the cluster center is dominated by the jet and its interaction with cold gas (especially for low-density jets). Consequently, a clear connection between cold gas and bubbles is expected. To this end, we analyse the cold gas morphology and kinematics by looking at internal alignment of the cold gas angular momentum. In the following we only consider cold cells with $T<2\times10^{4}\,\mathrm{K}$ or $\mathrm{SFR}>0$ which constitute the filamentary structures seen in our simulations. We define the angular momentum of those individual cells of cold gas, $\bm{l}_{\mathrm{fil},i}=m_i\bm{r}_i\btimes\bm{v}_i$ and calculate the total angular momentum of the cold gas via
\begin{align}
  \bm{L}_\mathrm{fil}=\sum_{i} m_i\bm{r}_i\btimes\bm{v}_i
\end{align}
where we measure all velocity and radius vectors with respect to the cluster center and only account for cold cells according to the criterion defined above. This enables us to compute the alignment statistics of individual cold cells with the total angular momentum by computing the vector product of the normalised cell's angular momentum $\bm{\hat{l}}_\mathrm{fil}$ with the unit total angular momentum $\bm{\hat{L}}_\mathrm{fil}$, i.e.\ $|\bm{\hat{L}}_\mathrm{fil}\bcdot\bm{\hat{l}}_\mathrm{fil}|$. Figure~\ref{fig:epsilonplot} shows projections of this quantity and PDFs where colors in the lower left corner of each image label individual distributions in the respective PDFs. The formation and presence of individual filamentary structures are transient phenomena on timescales of a 100s of Myr so that sufficient time-sampling is required. PDFs of a single model for different times are compared in the bottom panels, while different models at individual times are shown on the right. Distributions that peak at low values of $|\bm{\hat{L}}_\mathrm{fil}\bcdot\bm{\hat{l}}_\mathrm{fil}|$ are more random/filamentary than distributions peaking at higher values which are more discy.

In general, the models \texttt{Fiducial} and \texttt{Bondi} show very elongated filamentary structures extending out to $r\sim30$--$50\,\mathrm{kpc}$. We see a coherent kinematic structure along the filaments with a slowly varying angular momentum distribution. While there are also filamentary cold structures in the \texttt{Dense} model, they are confined to smaller radii. Magnetic fields in combination with radiative gas cooling cause the formation of these filamentary structures while purely hydrodynamic simulations shatter cooling clouds into small cloudlets as a result of Kelvin-Helmholtz instabilities \citep{Sparre2020,Mueller2021}, similar to the formation of long filamentary tails of jellyfish galaxies.

In the models \texttt{Fiducial} and \texttt{Bondi}, which enable extended filamentary structures, we only observe transient disc structures in the cold gas phase, which either foster star formation or get accreted onto the SMBH on short timescales of a few 100s of Myr. Importantly, in these models the disc is not continuously fed with gas of similar angular momentum, which is consistent with observations \citep{Russell2019}. By contrast, there are persistent and long-lived discy cold gas morphologies in the \texttt{Dense} and \texttt{HD} models. In particular, the \texttt{HD} model clearly shows a strong discy distribution of the cold gas phase throughout. In model \texttt{Dense}, there is a long-lived coherent discy structure at the cluster center, while at radii $r\gtrsim10\,\mathrm{kpc}$ we observe a more random distribution of filamentary structures. We see that in the presence of a cold-gas disc, the cold gas is much more confined to the center and less volume filling.

Magnetic fields allow for a more efficient coupling of cold and hot phases, by efficiently sharing momentum between these two phases through the magnetic pressure and tension forces \citep{Wang2021}. In addition, the increased magnetic field strength in the cooling gas leads to magnetic breaking, reducing angular momentum, which may limit disc formation \citep{Wang2020}. Here, magnetic fields appear to be necessary to transfer jet-induced angular momentum from the hot phase onto the cooling gaseous phase so that later accreting filaments condense into a central configuration with a different angular momentum distribution, thus precluding the formation of a sustained and massive disc. In general, jets induce turbulence in the central region of the cluster, which interferes with condensing gas. This in combination with the turbulence from the initial conditions affects the dynamics of the cold gas. While we do not follow the evolution of the cold gas to the H$\alpha$ and carbon-oxygen (CO) emitting phases, we speculate that the steady injection of momentum in the same direction of the high density jets appears to facilitate the formation of more discy structures that seem to be inconsistent with observed extended filamentary structures surrounding cD galaxies in clusters \citep{Russell2019}. On the other hand, the deflected low-density jets can lift up cold gas in their wakes and stir the cold gas in more isotropic directions.

\subsection{Influence of the magnetic field on AGN feedback}
\label{sec:magneticfield}
\label{sec:magneticfieldinfluence}
\begin{figure*}
\centering
\includegraphics[trim=0cm 0cm 0cm 0cm,clip=true, width=1.\textwidth]{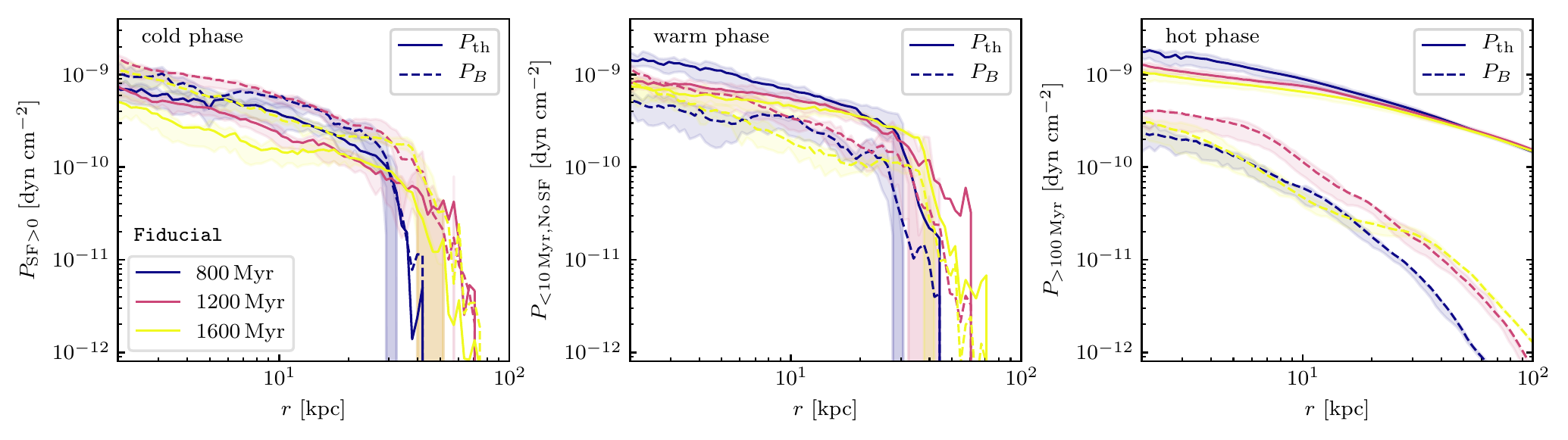}
\caption{Radial profiles of the thermal pressure (solid) and magnetic pressure (dashed). From left to right, we show the star forming cold gas ($P_{\mathrm{SF}>0}$), non-star forming, warm gas with $t_\mathrm{cool}<10\,\mathrm{Myr}$ ($P_{<10\,\mathrm{Myr},\mathrm{No}\,\mathrm{SF}}$) and hot gas with $t_\mathrm{cool}>100\,\mathrm{Myr}$ ($P_{>100\,\mathrm{Myr}}$). Radial profiles correspond to average profiles within $160\ \mathrm{Myr}$ and shaded areas denote the 10th to 90th percentiles. In the cold phase, the magnetic field becomes dynamically relevant because of a loss of thermal support as a result of radiative cooling.}
    \label{fig:radialprofilesmagneticpressure}
\end{figure*}

\begin{figure*}
	\centering
	\includegraphics[trim=0cm 0cm 0cm 0cm,clip=true, width=0.925\textwidth]{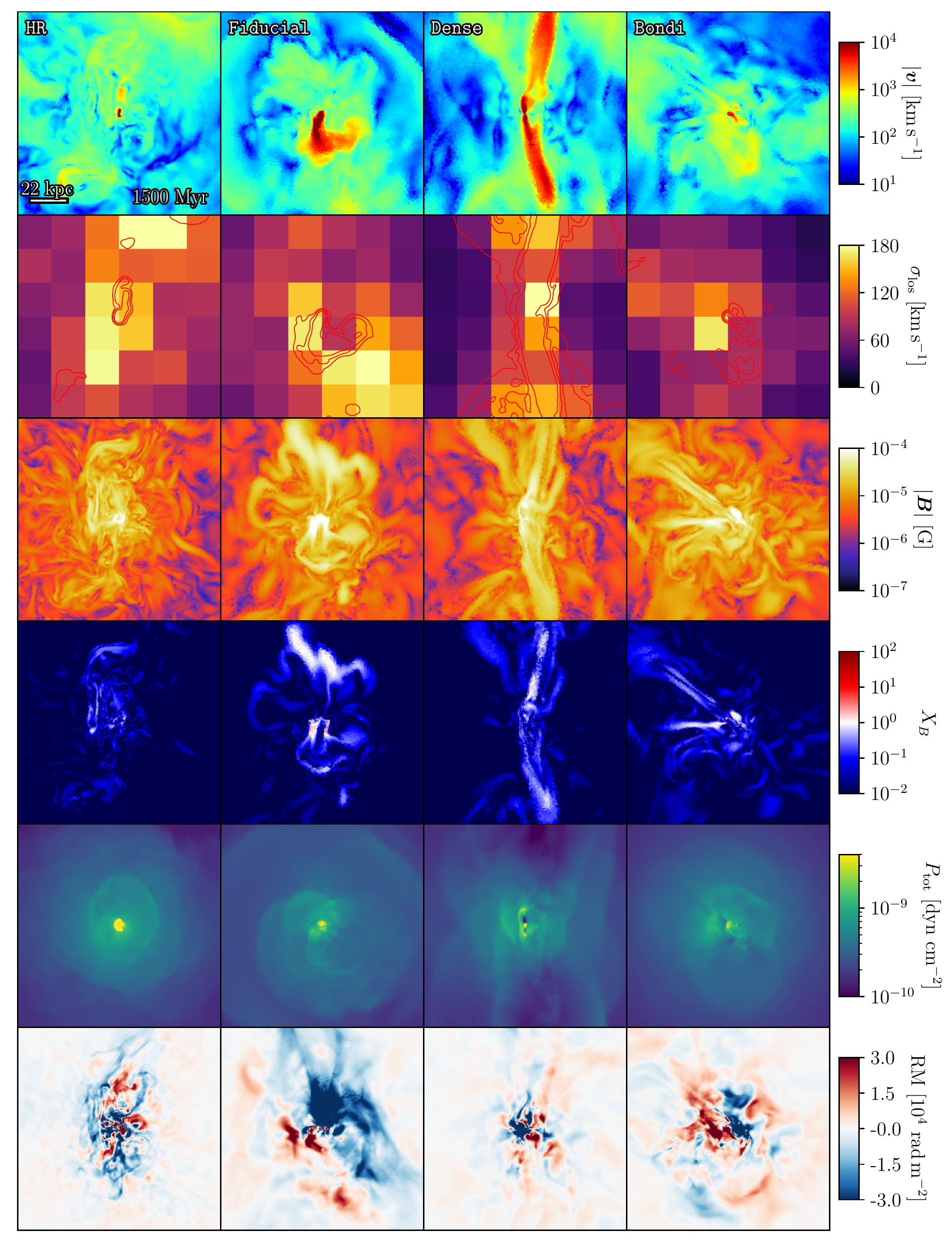}
	\caption{From top to bottom, we show slices of the velocity $|\vect{v}|$, X-ray weighted mean line-of-sight velocity dispersion, slices of the magnetic field, magnetic-to-thermal pressure ratio $X_B$, total pressure $P_\mathrm{tot}$ and Faraday rotation measure at 1500 Myr. Contours correspond to jet tracer values of $10^{-4}$, $10^{-3}$ and $10^{-2}$ in a midplane slice. Jet induced turbulence is visible as increased velocity dispersion. In our model \texttt{Dense}, high momentum-density jets propagate nearly unperturbed and cause conically enhanced velocities and velocity dispersion while low-momentum density jets in the three other models are more easily deflected, which results in more isotropic velocity fields. The magnetic field is strongly amplified in the wake of the jets due to increased cooling rates that are triggered by converging gas flows giving rise to substantial rotation measure values. }
	\label{fig:rm}
\end{figure*}

In Fig.~\ref{fig:radialprofilesmagneticpressure}, we show radial profiles of the thermal (solid) and magnetic (dashed) pressure in the model \texttt{Fiducial} at different times. From left to right, we show profiles in the cold phase ($\mathrm{SFR}>0$), warm phase ($t_\mathrm{cool}<10$~Myr, $\mathrm{SFR}=0$) and hot phase ($t_\mathrm{cool}>100$~Myr). Magnetic fields become progressively more dynamically relevant for colder phases to the point where the cold phase is dominated by magnetic pressure \citep{Wang2021}. Consequently, magnetic fields will play a significant role in the dynamics of filaments. Note, as our ISM model prevents cooling to observed temperatures that are orders of magnitudes lower than in our model, we expect thermal pressure losses due to additional cooling to increase the magnetic fields even further. Consequently, the implications for stability and dynamics are expected to be even more severe.

As discussed in \cite{Ehlert2021}, magnetic fields and velocity fields are inherently coupled via the induction equation and the equation of motion. It may therefore be instructive to look at the velocity field here.
In Fig.~\ref{fig:rm}, we show slices of the absolute velocity $|\bm{v}|$, line-of-sight velocity dispersion $\sigma_\mathrm{los}$ of the X-ray emitting gas ($k_\rmn{B}T<15\,\mathrm{keV}$) with contours of a jet tracers on a cut-plane through the cluster center, slices of the absolute magnetic field $|\bm{B}|$, the magnetic-to-thermal pressure ratio $X_B$, the total pressure $P_\mathrm{tot}$, and the Faraday rotation measure:
\begin{equation}
\mathrm{RM}=\frac{e^3}{2\pi m^2_\mathrm{e}c^4}\int_0^{L} n_\mathrm{e}\, \bm{B}\bcdot \mathrm{d}\bm{s},
\end{equation}
where the integral extends along the line of sight from the source to the observer, $e$ is the elementary charge and $m_\rmn{e}$ is the electron rest mass. The magnetic field and thermal electron density are results of our simulations.  For gas that is not forming stars, we directly use the thermal electron density as calculated in the cooling module of the code. For star-forming gas, we need to account for the subgrid-scale model of the ISM \citep{Springel2003} used in our simulations. This model implicitly assumes an unresolved multi-phase ISM consisting of a volume-filling warm phase and a neutral cold phase at $10^4$~K that dominates the gas mass. Hence, in order to calculate the contribution to the Faraday rotation measure of star-forming gas, we self-consistently calculate gas that our subgrid model assigns to the volume-filling warm phase as laid out in \citet{Springel2003} and assume that it is fully ionised. Note that we may still overestimate the electron number density as the AGN outflow may contain a molecular gas component that we do not account for here.

In Fig.~\ref{fig:rm}, the high-velocity jet is discernible especially in the \texttt{Dense} model, where the jet faces minimal deflection. However, corresponding line-of-sight velocities are small $\sigma_\mathrm{los}\sim120\,\mathrm{km}\,\mathrm{s}^{-1}$. In contrast, the \texttt{HR} and \texttt{Fiducial} models show multiple pixels with $\sigma_\mathrm{los}\gtrsim180\,\mathrm{km}\,\mathrm{s}^{-1}$ at this time. Here, larger fractions of the velocity are diverted into the line of sight. Due to the persistent jet directions in model \texttt{Dense}, the enhancement of the ICM velocity dispersion is limited to a cone region in jet direction. On the other hand, low-density jets induce more isotropic turbulence, which is reflected in the velocity dispersion maps where increased values coincide with the jet location. Propagating jets induce shocks in the ICM that are clearly discernible as regions of increased $P_\mathrm{tot}$ in Fig.~\ref{fig:rm}. While shocks in runs with low density jets appear spherical, the shocks in simulation \texttt{Dense} are more ellipsoidal and therefore mostly interact in the direction perpendicular to the direction of jet injection.

In our simulations, jets amplify the magnetic field in the wake to values of order 100~$\mu$G, as a result of converging gas flows that compress the gas \citep{Ehlert2021}. This increases the cooling rate so that the thermal gas quickly loses pressure support to the point where the magnetic pressure dominates over the thermal pressure, i.e., $X_B\gtrsim1$. In these regions, the magnetic tension force withstands the turbulent motions in the ICM, giving rise to comparably straight filaments (see Fig.~\ref{fig:rm}). Note that fluctuations in the total pressure appear due to jet-induced shocks. There is no obvious correlation between total pressure and the magnetic field structure, implying that these highly magnetized filaments are in approximate pressure equilibrium with the surrounding ICM.

By construction, our initial magnetic field strength $X_{B,\mathrm{ICM}}=0.0125$ yields bulk rotation measures that are in agreement with observations \citep{Clarke2004a,Murgia2011a}. However, in the regions of strong magnetic field, the rotation measure reaches values in excess of $\mathrm{RM}>3\times10^4~\rmn{rad~m}^{-2}$, which is an order of magnitude above observed values. However, the rotation measure morphology is a strong function of numerical resolution (bottom two panels of Fig.~\ref{fig:rm}). We see alternating regions with differing signs of the Faraday rotation measure on small angular scales, suggesting that beam smoothing effects and other observational uncertainties may bring our simulations closer to the observed results \citep[e.g.,][]{Newman2002,Johnson2020a}. Most importantly, the amplified magnetic field stays generally confined to the path of the bubble on simulated timescales. Thereby, low-density jets amplify magnetic field more isotropically compared to the unidirectional jets in our \texttt{Dense} model.

\section{Discussion}
\label{sec:discussion}
In this paper, we examined simulations of isolated CC galaxy clusters including radiative cooling, star formation and black-hole accretion-regulated feedback from AGN driven jets. We explore variations of different aspects of the simulations including the presence of magnetic fields, the SMBH accretion rate model, the density of the AGN driven jet as well as efficiency parameters. In all our simulations, radiative cooling is on average well-balanced by heating from AGN-driven jets, with at most a factor of a few higher instantaneous jet luminosities (compared to the halo cooling losses), but at times also orders of magnitude lower luminosities (Fig.~\ref{fig:cavityheatingcooling}).

\subsection{Self-regulated CC clusters}
In all explored model variations that include jets, the SFR in the cluster is suppressed by at least one order of magnitude compared to the no-feedback case (Fig.~\ref{fig:wangp}), yet the thermodynamic profiles remain characteristic for CC galaxy clusters, i.e.\ with high central densities at or exceeding $0.3$~cm$^{-3}$, central temperatures below $2.5$~keV, and entropies well below $10$~keV~cm$^2$ (Fig.~\ref{fig:radialprofilescc}). A \emph{gentle} mode of AGN feedback has also been demonstrated in simulations by \cite{Yang2016,Li2017,Meece2017,Bourne2021}, which they attribute to jet-induced shock heating and mixing. Furthermore, our jet feedback is unable to convert CC to non-CC clusters as also seen by \cite{Hahn2017,Chadayammuri2021}. In line with the resulting CC characteristics, a substantial amount of the central ICM gas remains in a state where cooling times are low compared to the free-fall time (Fig.~\ref{fig:radialprofilescooling}), cold gas is constantly present in the cluster (Fig.~\ref{fig:gasphases}), yet only a small, highly time-variable fraction of it directly feeds the SMBH in the center, causing the instantaneous jet luminosities to fluctuate substantially (Fig.~\ref{fig:jetpowerhistogram}). While the presence of cold gas is universal, we find its properties to be highly dependent on different modeling choices: our fiducial model of a magnetized ICM heated by light jets tends to create transient gas filaments with low circularity parameters (i.e.\ either on predominantly radial or uncorrelated orbits) extending several tens of kpc from the center, while two model variations create cold gas structures with large circularity parameter, i.e.\ coherent, disc-like rotation (Fig.~\ref{fig:epsilonplot}). This includes the simulation without magnetic fields, indicating that continuous feeding of gas with the same angular momentum is precluded by magnetic fields interconnecting the hot and cold phases. Even in the presence of magnetic fields, the model employing a dense jet produces coherently rotating cold gas discs, however, significantly more compact ones than the ones in the \texttt{HD} case. A plausible reason for this is the directionality of the jet  that clearly constrains the direction of resulting lobe structures (a proxy for the large-scale gas flow patterns) in the case of dense jets, but not in the case of light jets (Fig.~\ref{fig:cavityfittingexplanation}). This implies that in the case of low-density jets, centrally forming cold gas is dragged around by large-scale turbulent motions in the central region of the cluster, while dense jets facilitate a more coherent flow pattern that facilitates the buildup of a rotating disc. The resulting continuous presence of cold gas near the SMBH for dense jets leads to increased accretion rates as seen in Figure \ref{fig:wangp}.

We fitted ellipses to X-ray images of our simulations to more easily connect our simulations to observations of X-ray cavities. Our cavity luminosities are in the range $10^{44}\,\mathrm{erg}\,\mathrm{s}^{-1}\lesssim L_\mathrm{cav}\lesssim 3\times10^{45}\,\mathrm{erg}\,\mathrm{s}^{-1}$ across simulations, which is in general agreement with the observed total bubble luminosity in Perseus \citep{Birzan2004,Rafferty2006,Diehl2008}. The corresponding spread in cavity luminosities of four orders of magnitude agrees with the variance observed when looking at cluster samples at similar ICM luminosities \citep[see Fig.~6 in][]{Rafferty2006}.

We found a successful solution for self-regulation in a Perseus-like cluster based on physical principles. However, the shallower potential of groups and smaller clusters is expected to show a different coupling efficiency of the AGN and the ICM as suggested by \cite{Prasad2020}. On the other hand, in the Phoenix cluster, one of the most massive clusters observed, AGN feedback appears to be too inefficient to halt cooling \citep{McDonald2019}. Therefore more simulations across the cluster mass range in a cosmological setting are required to demonstrate that the presented models are able to successfully self-regulate CC clusters. Nevertheless, our work opens up an avenue to simulate realistic CC clusters in cosmological zoom-in simulations and to finally answer the question about the origin of the bimodality of CC and non-CC systems.

\subsection{Accretion models and jet propagation direction}

For our simulated scales the exact accretion model has limited relevance. Both Bondi and chaotic cold accretion are very sensitive to dense, cold gas, which makes up the majority of gas accreted in our runs (see Fig.~\ref{fig:gasphases}). The additional continuous accretion in the Bondi model leads to small outbursts with limited influence on the cluster. In agreement with simulations by \cite{Meece2017}, we find that the exact triggering mechanism is secondary for runs with sufficiently high resolution. For the cold accretion model, CCs stay intact independent of probed choices for $\epsilon$ and $\eta$. However, our run with $\epsilon=1,\,\eta=0.0001$ grows the SMBH to $M>10^{11}\,\mathrm{M}_\odot$ within $2\,\mathrm{Gyr}$ (see Fig.~\ref{fig:wangpmodelparameters}), while observed most massive SMBHs have smaller masses: $6.6\times10^{10}\,\mathrm{M}_\odot$ \citep[TON 618;][]{Shemmer2004}, $5.1\times10^{10}\,\mathrm{M}_\odot$ \citep[MS0735-BCG;][]{Dullo2019}, $4\times10^{10}\,\mathrm{M}_\odot$ \citep[Holmberg 15A;][]{Mehrgan2019}. Observational biases should favor more massive SMBHs rather than smaller ones. In addition, \cite{King2016} shows that active SMBHs can grow to a maximum mass $M_\mathrm{max}\simeq5\times10^{10}\,\mathrm{M}_\odot$ for typical parameters and only reach $2\times10^{11}\,\mathrm{M}_\odot$ in rather extreme cases over a Hubble time. Hence, we conclude that this parameter combination is unlikely to be realised in nature.

We parametrise our accretion models by two variables, the accretion efficiency $\epsilon$ in the cold accretion model and the accretion-to-jet power conversion efficiency, $\eta$. We find that the total jet efficiency, i.e.\ the product of the two, $\epsilon\,\eta$ is most important for describing the properties of the ICM rather than varying both parameters individually (see Fig.~\ref{fig:wangpmodelparameters} and Appendix~\ref{app:modelparameters} for the detailed analysis). In particular, we find that the jet efficiency $\epsilon\,\eta$ determines the level of intermittency in our runs. More efficient jets (higher $\epsilon\,\eta$) are able to temporarily push and/or drag cold gas out from the central regions. Lower efficiencies force the jet to be active at all times, limiting jet powers to the maximum while extra cooling is converted into star formation. On the other hand, more energetic outbursts from high-power jets can increase cooling times on larger scales, which can possibly reduce the amount of newly collapsing gas and stabilize the atmosphere more efficiently. However, we note that generally cooling gas with $t_\mathrm{cool}<10\,\mathrm{Myr}$ and star formation are always present in all simulations. The large variety of efficiency parameters (with $\eta=1$) of self-regulated feedback models used by other research groups generally confirm our findings, which range from $\epsilon=6\times10^{-5}$ \citep{Prasad2015}, $\epsilon=10^{-4}$ \citep{Prasad2020}, $\epsilon=5\times10^{-4}$ \citep{Prasad2018}, $\epsilon=10^{-3}$ \citep{Li2014b,Meece2017}, $\epsilon=5\times10^{-3}$ \citep{Wang2019} to $\epsilon=10^{-2}$ \citep{Li2015}. While these models vary in exact jet powers and SFRs, we consider the similarities of the self-regulated state reassuring considering the vast spectrum of codes, jet models, ISM implementations and resolutions used. More detailed analysis is required to constrain specific parameters.

Our clusters are in a state of condensation with minimum cooling-to-free fall time ratios $\min(t_\mathrm{cool}/t_\mathrm{ff})<10$, with some a fraction of the ICM reaching $\min(t_\mathrm{cool}/t_\mathrm{ff})<1$ (see Fig.~\ref{fig:radialprofilescooling} middle panel). Previous simulations generally report instability for gas with $t_\mathrm{cool}/t_\mathrm{ff}<10$ \citep{Sharma2012,Gaspari2012,Gaspari2014,Choudhury2019a}. Observations of cold filaments in clusters find similar values \citep{Voit2015c} but also somewhat larger values in the range $10\lesssim t_\mathrm{cool}/t_\mathrm{ff}\lesssim25$ \citep{Pulido2018,Olivares2019}. The low end of the entropy distribution at a certain radius may explain high values in clearly thermally unstable clusters \citep{Voit2021}.

As discussed in Section \ref{sec:jet}, the random nature of (chaotic cold) accretion limits the expected jet precesssion considerably. However, we find that the exact jet direction is extremely important for heavy jets because their high-momentum density implies a fast transport to larger radii and minimizes the lateral heating rate so that the gas quickly becomes thermally unstable and feeds the SMBH perpendicular to the jet direction. Varying the jet direction should have a smaller impact on the self-regulation in the light-jet models because their propagation direction is determined by deflection events off of cold gas filaments and clouds, leading to more isotropic heating in the central regions owing to the increased lateral momentum deposition.

\subsection{H$\alpha$ and CO filaments and the role of non-thermal components}

While observed filaments reach down to smaller temperatures than simulated here, we argue that comparing our results to observations is nevertheless instructive. Observed filaments in \halpha and CO show a large distribution of morphologies, ranging from disc-dominated to filamentary structures that can extend over several kpc in length \citep[e.g.,][]{Russell2016b,Gendron-Marsolais2018}. Most cold gas and filaments are located below or near bubbles (see Fig.~\ref{fig:overviewhighres}). In addition, smooth velocity gradients along the filaments support the idea that buoyantly rising bubbles lift central cold gas and/or lifted gas becomes thermally unstable and cools \citep{Russell2019}. This agrees with our results on the kinematics and morphology of cold phase filaments (see Fig.~\ref{fig:epsilonplot}). \cite{Beckmann2019a} analyse cold gas in their hydrodynamical simulations and find that filaments are easily shattered by the jet leading to overly clumpy morphologies compared to observations. While we defer a detailed analysis to future work, we also notice clumps in our simulations that are not directly connected in filaments.

In general, the exact details of the feedback loop sensitively depend on the cooling processes via the thermal instability which is in turn sensitive to resolution effects \citep{Martizzi2019}, limiting the scope of detailed quantitative predictions from our simulations. However, our higher resolution runs confirm the  findings discussed throughout the paper (see Appendix \ref{app:resolution} for details). In addition, a detailed analysis of resolution dependencies for the jet can be found in \weinberger

Magnetic fields lead to a strong coupling between cold and hot gas phases \citep{Wang2021}, by sharing momentum between these two phases through the magnetic pressure and tension forces. As a result, the cold phase adopts a more filamentary morphology \citep{Sparre2020} as magnetic draping suppresses Kelvin-Helmholtz instabilities \citep{Ruszkowski2007,Dursi2007,Dursi2008,Ehlert2018}. As such, magnetic fields preclude the formation of long-lived cold discs in the first place and do not have to disrupt an already formed disc. In addition, magnetic fields provide substantial pressure support to the cold gas.

Analogously, cosmic ray protons prone to the streaming instability can potentially provide stability to H$\alpha$ filaments and provide heating to power the emission \citep{Ruszkowski2018}. In addition, Alfv\'en heating from streaming cosmic ray protons may be the dominant heating mechanism in CC clusters with profound implications on the dynamics of the cluster and the resulting feedback cycle \citep[e.g.,][]{Guo2008,Pfrommer2013,Jacob2016b,Ruszkowski2017a,Ehlert2018,Wang2020}. We defer their inclusion to future work.

\section{Conclusions}
\label{sec:conclusions}
Jet feedback is able to stabilize CC galaxy clusters against thermal collapse. We employ MHD simulations to study AGN feedback in an idealised turbulent Perseus-like cluster. Our findings can be summarized as follows:
\begin{itemize}
	\item Independent of the accretion model (Bondi vs.\ chaotic cold accretion), probed accretion efficiency, magnetisation of the cluster and jet density, the cluster settles into a state of self-regulation after $\sim500\,\mathrm{Myr}$ (see Figs.~\ref{fig:wangp} and \ref{fig:wangpmodelparameters}) with density, entropy and cooling time consistent with observed CC clusters (see Fig.~\ref{fig:radialprofilescc}).
	\item More efficient jet feedback leads to more intermittent jet power and star formation (see Figs.~\ref{fig:wangp} and \ref{fig:wangpmodelparameters}).
	\item Our fiducial low-density jets are easily deflected by cold gas, which leads to more isotropic turbulence injection and bubble distributions (see Fig.~\ref{fig:overviewhighres}). The dense jets form bubble distributions almost exclusively in the jet direction (see Fig.~\ref{fig:cavityfittingexplanation}). Here, cooling gas is continuously funneled onto the SMBH perpendicular to the jet direction, which leads to quick accretion from a discy distribution that is much more confined to the cluster center ($r\lesssim10\,\mathrm{Myr}$, see Fig.~\ref{fig:epsilonplot}).
	\item Our purely hydrodynamic run forms a massive disc ($r\gtrsim10\,\mathrm{kpc}$) for $t\gtrsim200\,\mathrm{Myr}$ (see Fig.~\ref{fig:epsilonplot}) which leads to unrealistically high SFRs (see Fig.~\ref{fig:wangp}). Transient disc formation that is not long-lived in our MHD runs supports the idea that magnetic fields anchored in the hot phase redistribute angular momentum with the cooling gaseous phase so that filaments accreting later settle into a central configuration with a different angular momentum distribution, thus precluding the formation of a sustained and massive disc.
	\item In broad agreement with observations, a plethora of cold gas morphologies ranging from discy to very extended filamentary structures is observed across our MHD runs but morphology and extent ($r<40\,\mathrm{kpc}$) varies significantly with time in individual runs (see Fig.~\ref{fig:epsilonplot}).
	\item Our inferred luminosities from cavity size measurements correspond to averaged jet powers that are therefore insensitive to periods of short and low-luminosity jet injection. Our imposed temperature threshold for cold accretion leads to $\sim50\,\mathrm{Myr}$ long breaks in jet injection in comparison to the continuous accretion condition prescribed in the Bondi model. However, the long life-time of cavities of $>100\,\mathrm{Myr}$ since injection leads to a comparable presence of cavities for both models (see Fig.~\ref{fig:cavityheatingcooling}).
    \item The magnetic field is strongly amplified in the wake of the jets due to the increased cooling rates that are triggered by converging gas flows, giving rise to substantial rotation measure values (see Fig.~\ref{fig:rm}).

\end{itemize}
AGN feedback has been established as the main heating source that allows CCs to reach a state of self-regulation at the observed levels. In order to rule out any specific accretion and jet models, secondary observables such as bubble and cold gas morphologies are needed.

In a next step, accounting for the cosmological assembly of a galaxy cluster is crucial for obtaining a more realistic environment and evolution. We find that light AGN jets are required to obtain the observed extended filaments with an averaged isotropic distribution of the kinematics of filaments. Numerically converged heating rates in the light-jet model imply a minimum resolution of $0.6\,\mathrm{kpc}$ in the jet launching region (\weinberger). We caution that further zooming into the acretion region could reveal distinct smal-scale differences for the accretion model (Bondi vs.\ chaotic cold accretion). In addition, future work will model the cold gas more accurately so that we may disentangle different AGN models by means of the observed distributions of \halpha and CO filaments.

\section*{Acknowledgements}
KE and CP acknowledge support by the European Research Council under ERC-CoG grant CRAGSMAN-646955 and ERC-AdG grant PICOGAL-101019746.
This work was supported by the Natural Sciences and Engineering Research Council of Canada (NSERC), funding reference \#CITA 490888-16. The authors gratefully acknowledge the Gauss Centre for Supercomputing e.V. (www.gauss-centre.eu) for funding this project by providing computing time on the GCS Supercomputer SuperMUC-NG at Leibniz Supercomputing Centre (www.lrz.de).

\section*{Data availability}
The data underlying this article will be shared on reasonable request to the corresponding author.


\bibliographystyle{mnras}

\bibliography{library}



\appendix

\section{Model parameters}
\label{app:modelparameters}
\begin{figure*}
	\centering
	\includegraphics[trim=0cm 0cm 0cm 0cm,clip=true, width=0.79\textwidth]{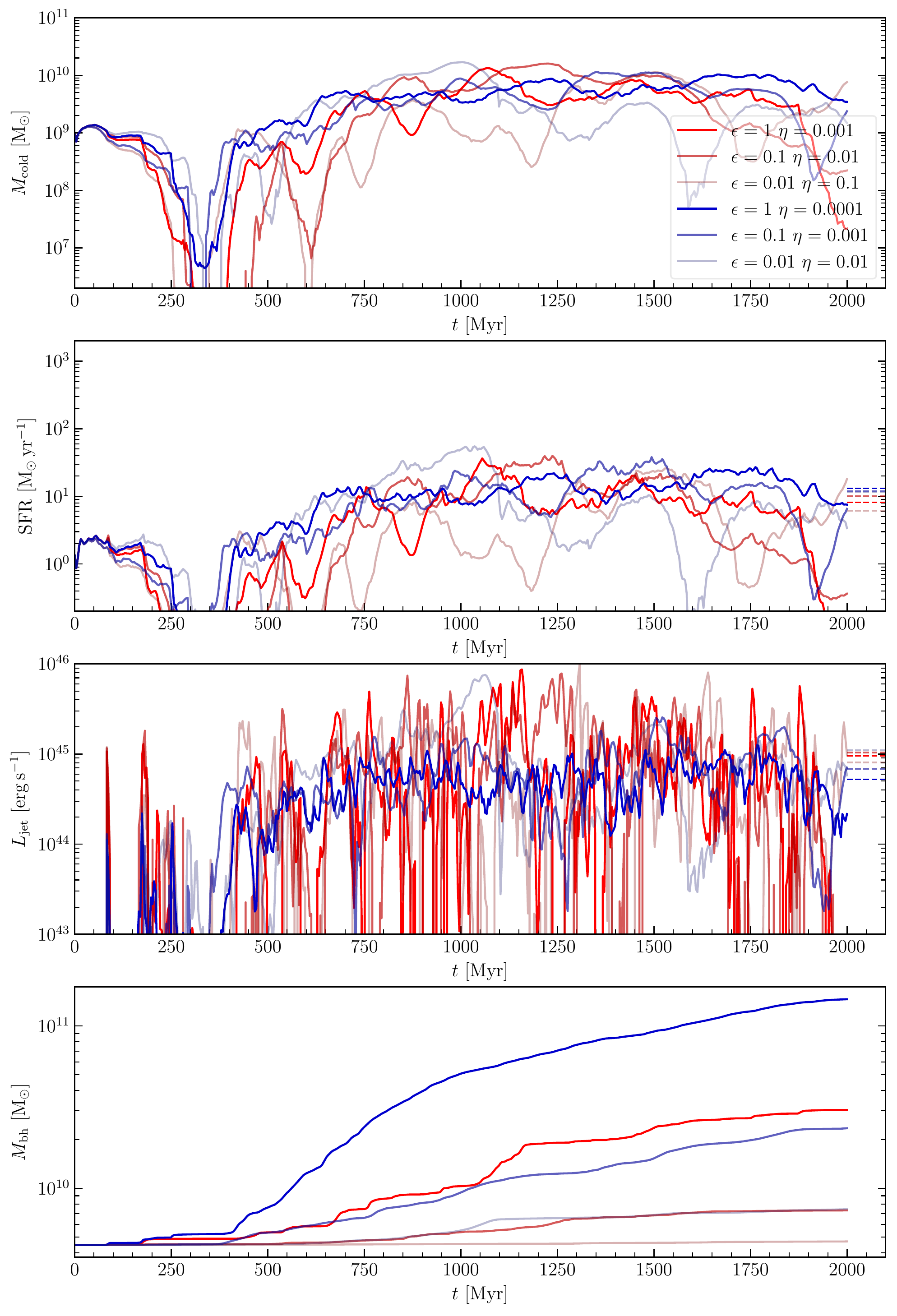}
	\caption{Same as in Fig.~\ref{fig:wangp} for runs with varying parameters of the cold accretion model. Analogously, all models attain to a state of self-regulation with very similar average SFRs ($\mathrm{SFR}\sim10\,\mathrm{M}_{\odot}\,\mathrm{yr}^{-1}$) and jet powers ($L_\mathrm{jet}\sim10^{45}\,\mathrm{erg}\,\mathrm{s}^{-1}$). Red and blue color shadings correspond to runs with $\epsilon\times\eta=0.001$ and $\epsilon\times\eta=0.0001$, respectively. Runs with lower coupling efficiencies show almost constant jet powers and SFRs while higher efficiencies cause star formation and jet powers to become more intermittent.}
	\label{fig:wangpmodelparameters}
\end{figure*}
\begin{figure*}
	\centering
	\includegraphics[trim=0cm 0cm 0cm 0cm,clip=true, width=0.79\textwidth]{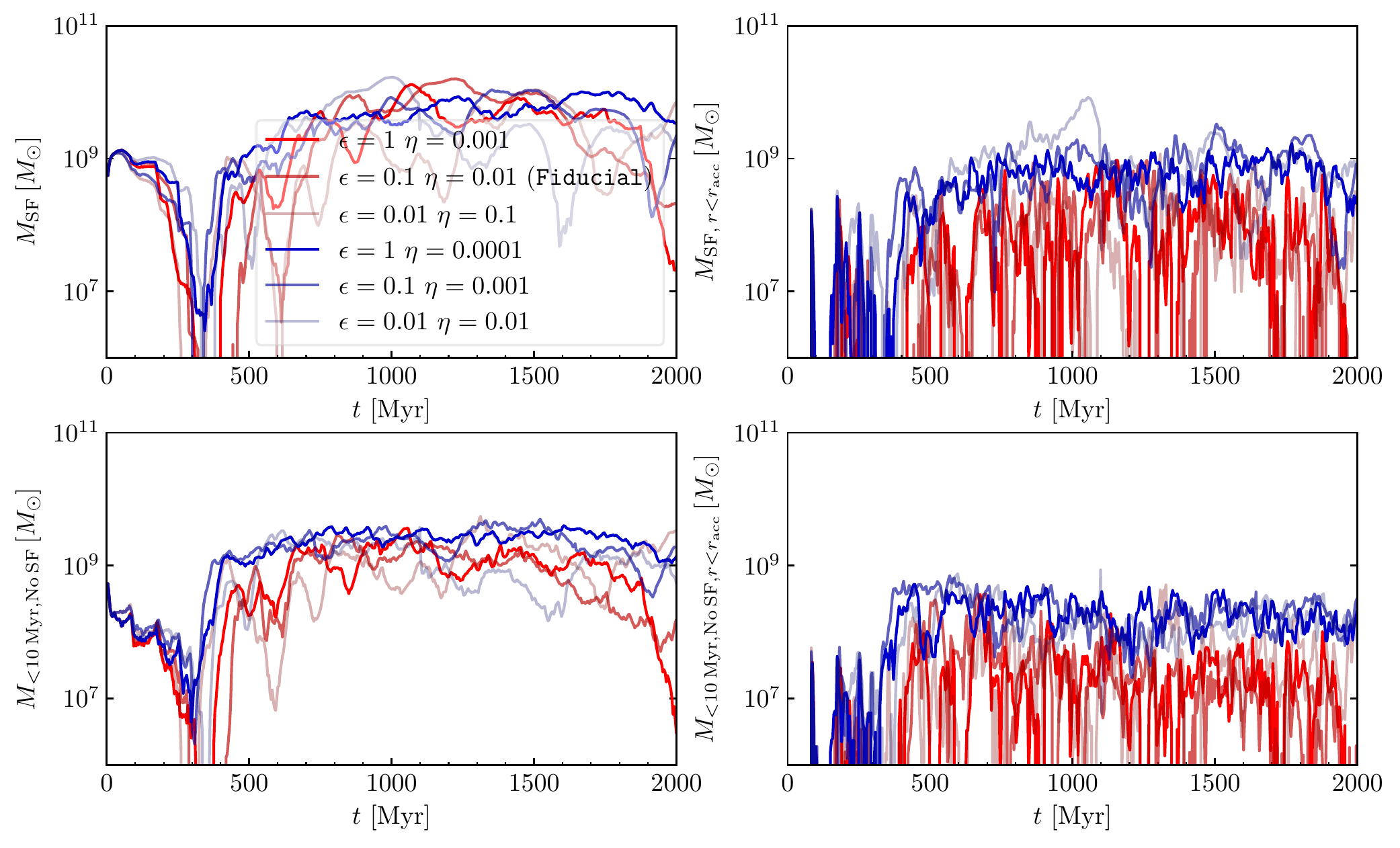}
	\caption{Same as in Fig.~\ref{fig:gasphases} for runs with varying parameters of the cold accretion model. Red and blue color shadings correspond to runs with $\epsilon\times\eta=0.001$ and $\epsilon\times\eta=0.0001$, respectively. Runs with lower efficiencies have a constant supply of cold gas that keeps powering the jets. More intermittent fluctuations are seen in the high efficiency runs. However, total cold and star forming gas masses stay within an order of magnitude.}
	\label{fig:gasphasesvaryparameters}
\end{figure*}

To assess the relevance of parameter choices governing jet efficiency, we varied both $\epsilon$ and $\eta$ independently. However, we find that only their product $\epsilon\times\eta$, which we refer to as \textit{jet efficiency}, gives significant differences in the cluster evolution. In Fig.~\ref{fig:wangpmodelparameters}, we report on the evolution of the cold gas mass (with $T<10^6\,\mathrm{K}$), SFR, jet luminosity $L_\mathrm{jet}$ and SMBH mass $M_\mathrm{bh}$ for runs with $\epsilon\times\eta=0.001$ (red) and $\epsilon\times\eta=0.0001$ (blue).

Runs with increased jet efficiency (reddish colours) in general show greater intermittency in star formation and jet power. More efficient jets pump more energy into the surrounding medium and halt cooling on longer timescales. Therefore, lower SFRs are observed. In addition, an increased injection of momentum forces cold gas on larger orbits around the center so that the accretion region is temporarily devoid of cold gas even though it remains present in the cluster  (see Fig.~\ref{fig:gasphasesvaryparameters}).

On the other hand, lower jet efficiencies (bluish colours) lead to the formation of cold gas that is more closely tied to the center and is depleted at a constant \textit{maximum} rate. Jet feedback is more continuous and less time-variable, while the high efficiency runs show higher burst powers and intermittent states of inactivity. Only the run with $\epsilon=0.01,\,\eta=0.01$, shows any significant variance after $t>500\,\mathrm{Myr}$ as a massive disc forms at $t\sim1\,\mathrm{Gyr}$ that is funneled into the accretion region on short timescales. Cooling is more significantly halted in the overheated cluster compared to its analogues at the same jet efficiency.

\section{Resolution}
\label{app:resolution}
\begin{figure}
	\centering
	\includegraphics[trim=0cm 0cm 0cm 0cm,clip=true, width=0.5\textwidth]{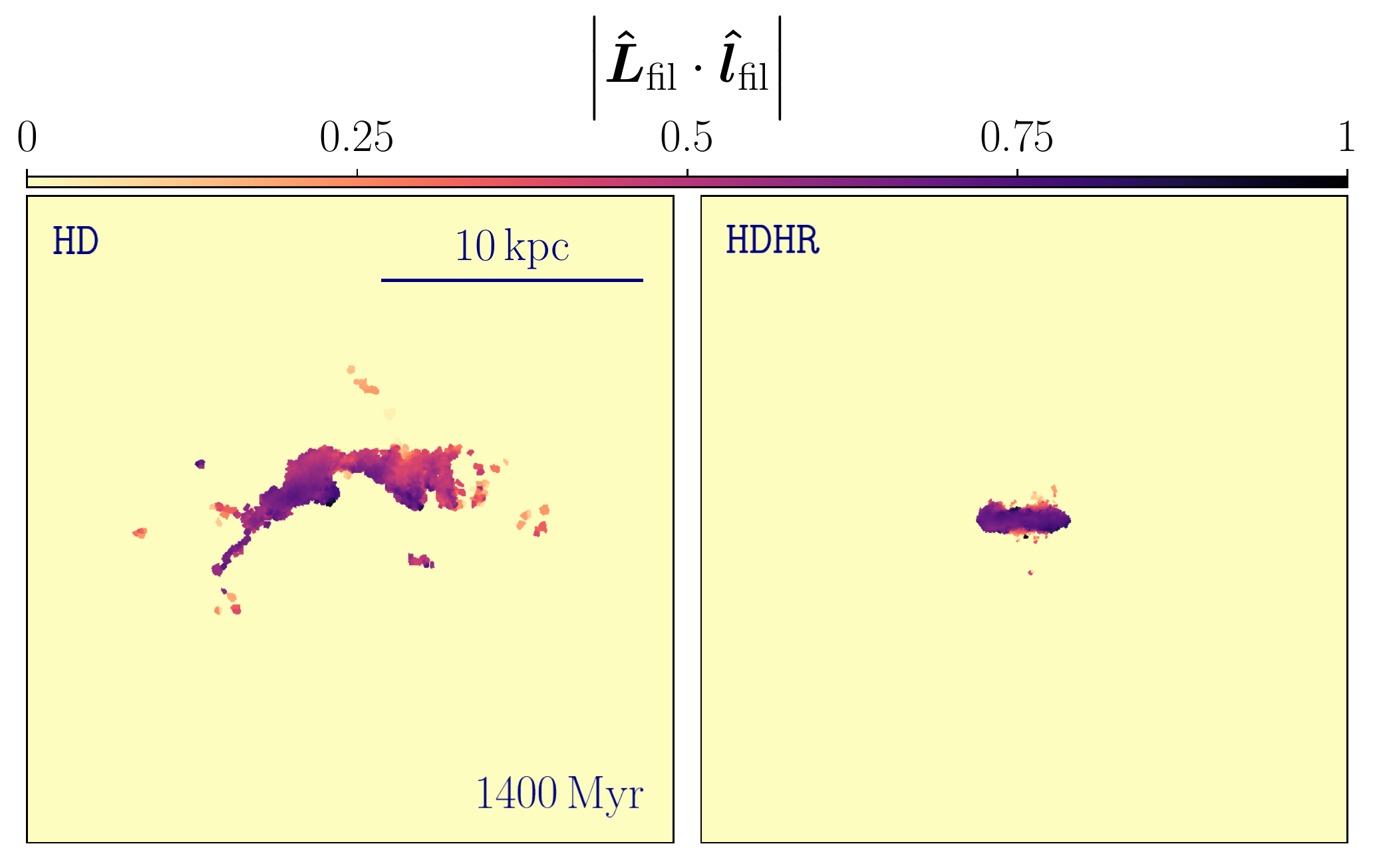}
	\caption{Same as in Fig. \ref{fig:epsilonplot} for \texttt{HD} and \texttt{HDHR} at 1400 Myr. Both hydrodynamical runs form a strong disc at the end of the run. In \texttt{HDHR} the disc is funneled into the SMBH completely depleting the entire cold gas reservoir in the cluster. }
	\label{fig:HDvsHDHR}
\end{figure}

\begin{figure*}
	\centering
	\includegraphics[trim=0cm 0cm 0cm 0cm,clip=true, width=0.95\textwidth]{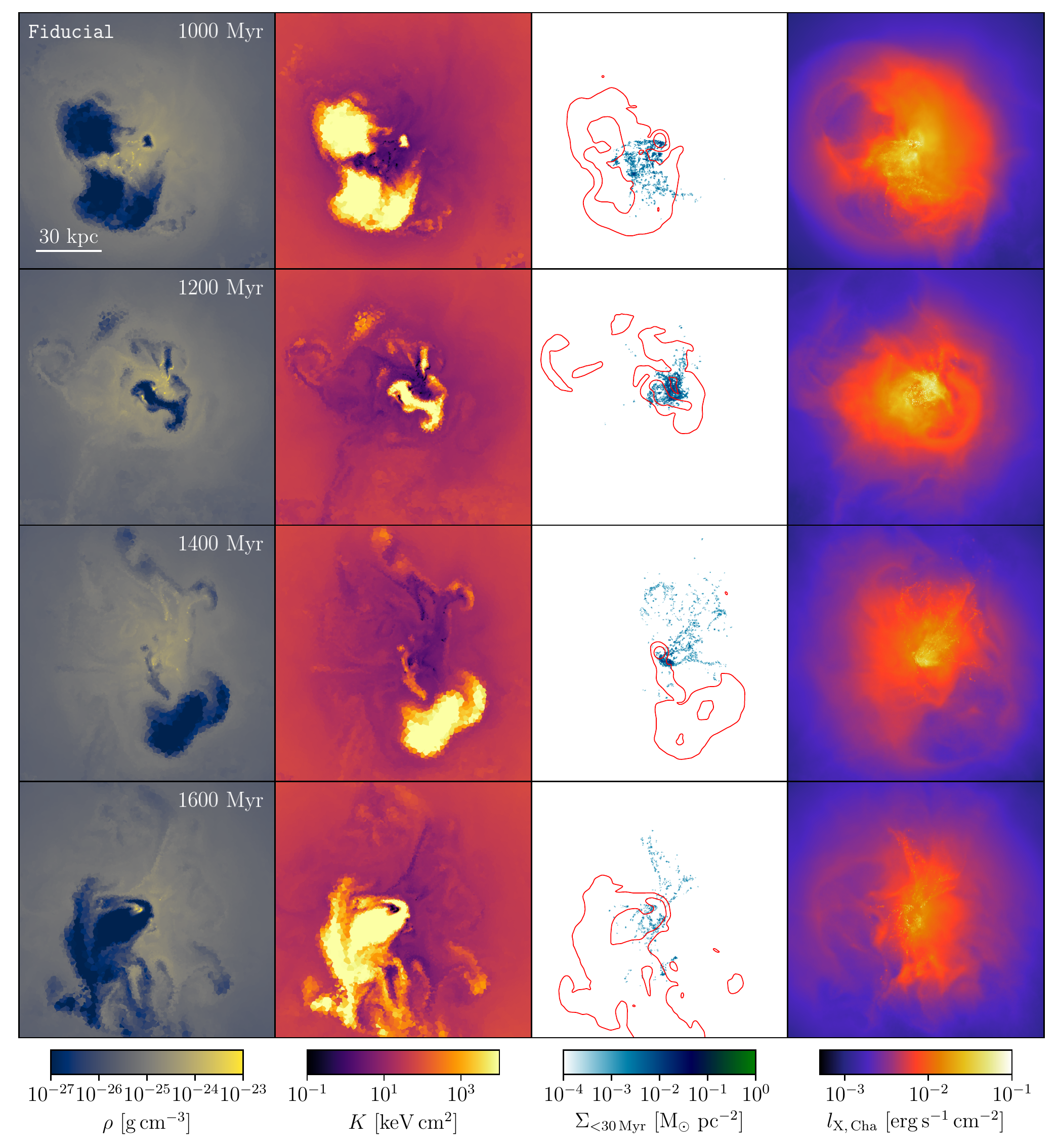}
	\caption{Same as in Fig.~\ref{fig:overviewhighres} for the same run at fiducial resolution \texttt{Fiducial}. The main features are independent of resolution.}
	\label{fig:overviewfid}
\end{figure*}

\begin{figure*}
	\centering
	\includegraphics[trim=0cm 0cm 0cm 0cm,clip=true, width=0.79\textwidth]{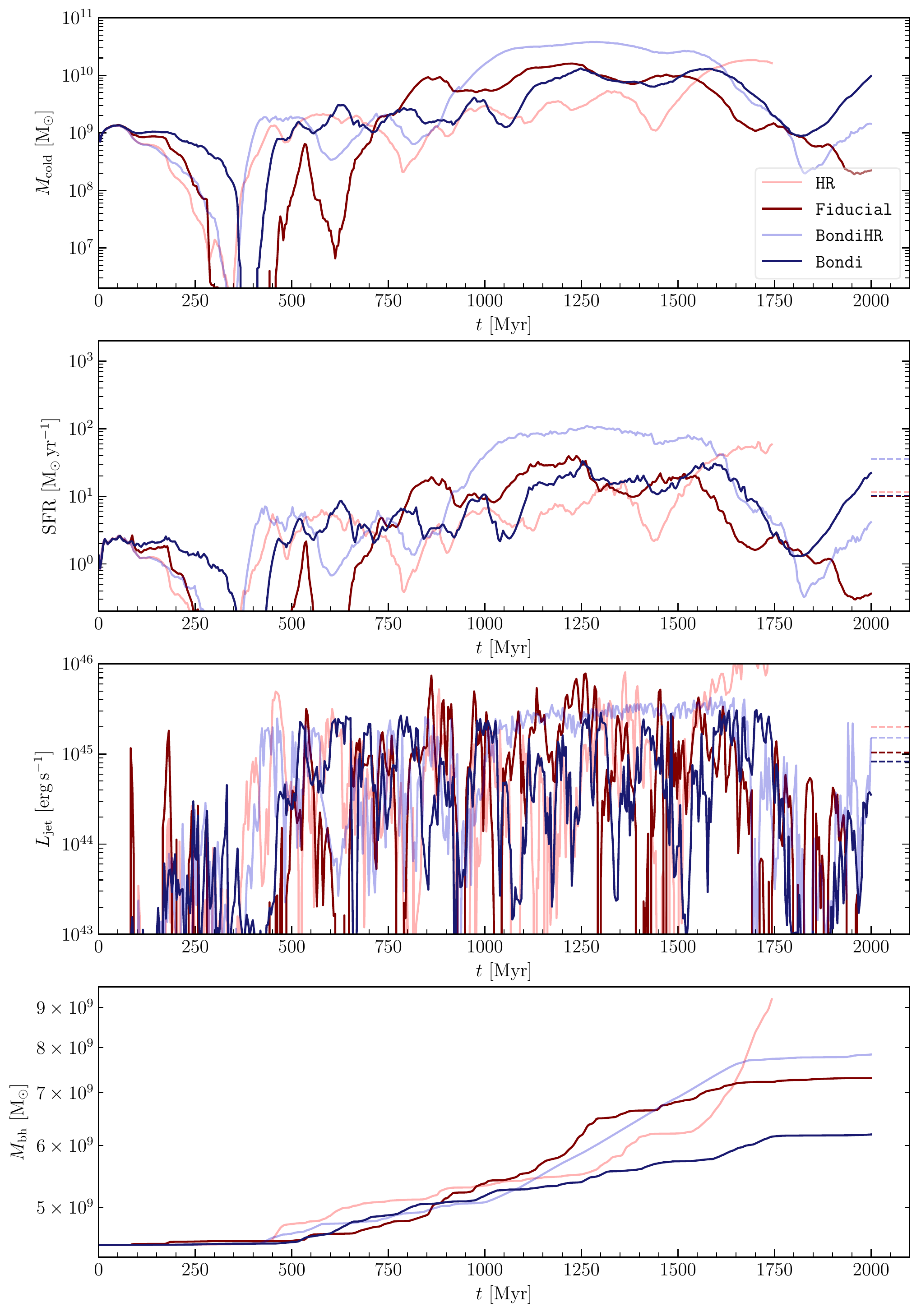}
	\caption{Same as in Fig.~\ref{fig:wangp} for fiducial cold accretion and Bondi runs at varying numerical resolution. The cold accretion run gives very similar results at both resolutions. During the high resolution Bondi run a transient disc is formed with high SFRs ($\mathrm{SFR}\sim100\,\mathrm{M}_\odot\,\mathrm{yr}^{-1}$). The reason for its formation should probably be attributed to a chance event rather than the difference in resolution as varying parameters may also lead to the formation of transient discs. }
	\label{fig:wangpresolution}
\end{figure*}

In agreement with our findings for the \texttt{HD} model, the high resolution analogue \texttt{HDHR} also forms a massive disc at $t\sim1000\,\mathrm{Myr}$, which is accreted within $\Delta t\sim800\,\mathrm{Myr}$ (see Fig.~\ref{fig:HDvsHDHR}). Hence, this demonstrates numerical convergence of the main properties of our hydrodynamic simulations. Throughout the simulation \texttt{HDHR}, the jet injects energy with $L_\mathrm{jet}>10^{45}~\rmn{erg~s}^{-1}$, which is significantly more than in any other run. This leads to an overheated core so that no further cooling is triggered until $t=2\,\mathrm{Gyr}$. We attribute this behavior to incomplete coupling with the hot and cold phases. As established earlier, magnetic fields are required to inhibit uncontrolled disc formation, we therefore omit the run from the following analysis.

In Fig.~\ref{fig:overviewfid}, we show our \texttt{Fiducial} model at the fiducial resolution. Compared to the high resolution analogue shown in Fig.~\ref{fig:overviewfid}, the general features are retained. Low-density bubbles rise that are deflected by cold gas in the ICM. Cavities are clearly discernible in X-ray emissivity.

In Fig.~\ref{fig:wangpresolution}, we show the cold gas mass ($M_\mathrm{cold}$ with $T<10^6\,\mathrm{K}$), SFR, jet luminosity ($L_\mathrm{jet}$) and SMBH mass ($M_\mathrm{bh}$). The evolution of the cold gas mass, star formation and jet power at high resolution are in general agreement with the low resolution counterparts. However, \texttt{BondiHR} forms a disc at $t\sim1,\mathrm{Gyr}$, which has vanished $500\,\mathrm{Myr}$ later. The related increased star formation and elevated jet power due to constant feeding leads to a somewhat distinct development. However, continued cooling quickly restores its earlier state of intermittent jet power and star formation as observed in \texttt{Bondi}. Similarly, the model \texttt{HR} shows a sudden increase in jet power at $t\sim1750\,\mathrm{Myr}$ probably due to accumulation of cold gas in the accretion region, which leads to a sudden spike in SMBH growth. We conclude that simulation-to-simulation variance appears more significant than resolution effects.

\bsp
\label{lastpage}
\end{document}